\documentclass[letterpaper,journal]{IEEEtran}
\usepackage[colorlinks,urlcolor=blue,linkcolor=blue,citecolor=blue]{hyperref}
 
\usepackage{color,array}
\usepackage{makecell}   
\usepackage{graphicx}
\usepackage{subcaption}
\usepackage{float} 
\usepackage{graphicx}
\usepackage{amsmath}
\usepackage{amssymb}  %

\usepackage{amsmath,amsfonts}
\usepackage{algorithm2e}
\usepackage{array}
\usepackage{textcomp}
\usepackage{stfloats}
\usepackage{url}
\usepackage{verbatim}
\usepackage{graphicx}
\usepackage{cite}
\usepackage{multirow}
\usepackage{booktabs} 
\begin{document}

\title{TIER: Trajectory-Invariant Explanation Regularization for Membership Privacy}

\author{{Varun Sharma , varun.sharma@stengg.com} {\\ Kar Wai Fok , fok.karwai@stengg.com\\}  { Vrizlynn L. L. Thing , vriz@ieee.org \\ ST Engineering, Singapore }}

\markboth{Journal of \LaTeX\ Class Files,~Vol.~14, No.~8, August~2021}%
{Shell \MakeLowercase{\textit{et al.}}: A Sample Article Using IEEEtran.cls for IEEE Journals}


\maketitle

\begin{abstract}
Explainability is central to building trustworthy AI, yet explanation interfaces can inadvertently provide adversaries with an expanded privacy-related attack surfaces. Recent studies show that advanced membership-inference attacks succeed by exploiting confidence‑drop trajectories, induced through attribution-guided perturbations, as discriminative features, rather than directly using confidence scores or explanation vectors. Existing defenses against membership inference fail to directly mitigate such explanation‑driven attacks. In this work, we investigate whether, during training, a model’s own gradients can be leveraged as defense signals against such attacks, thereby aligning explanation profiles between members and non‑members. To this end, we propose a \textbf{\textit{Trajectory-Invariant Explanation Regularization (TIER)}} defense that penalizes erratic fluctuations in confidence drops simulated  through gradient‑guided perturbations and simultaneously minimizes the distributional shifts via KL-divergence. Unlike conventional adversarial training, which emphasizes \textit{label robustness}, our approach targets \textit{explanation robustness} by enforcing self‑consistency through KL‑divergence and reducing the variance of confidence drops between members \& non-members. Extensive experiments confirm that our method effectively mitigates these attacks, delivering privacy protection while maintaining model utility and explanation fidelity.
\end{abstract}


\section{Introduction}
\IEEEPARstart{M}{embership} inference attacks compromise privacy by allowing adversaries to probe machine learning models and infer whether particular records were included in training. Unlike conventional breaches that involve direct file theft, these attacks exploit subtle statistical traces embedded in the model’s behavior. When such traces reveal the presence of sensitive data—such as medical images, payroll records, or personal conversations—the confidentiality of that information is undermined raising serious legal and ethical concerns.
Such traces often stem from overfitting: \textit{when a model memorizes training data, it tends to assign unusually high confidence or low loss to familiar inputs, leaving detectable patterns that adversaries can exploit.} Among the various strategies adversaries employ, one common approach is to construct shadow models on publicly available data and train attack classifiers to distinguish members from non‑members based on the shadow model’s confidence vectors.

However, recent attacks have advanced beyond using confidence score vectors or loss trajectories as features, instead leveraging explanatory maps to exploit privacy leakage via guided perturbations. In particular, \cite{10646875} demonstrate that this leakage intensifies when attributions are used to steer input modifications, since perturbing semantically important features identified by attribution maps produces a larger confidence drop for member samples than for non-members. By applying MoRF (\textit{Most Relevant Features, i.e., masking pixels ranked highest in importance}) and LeRF (\textit{Least Relevant Features, i.e., masking pixels ranked lowest in importance}) with distribution‑preserving operators, the authors demonstrate that attribution‑guided perturbations produce distinctive confidence‑drop trajectories, which serve as the primary features driving privacy leakage. They further show that existing defense approaches are either ineffective against attribution-guided perturbation attacks or impose severe utility constraints on the target model, thereby rendering it impractical.

Building on these observations, we hypothesize that \textit{the very same attack features can be transformed into defense signals by exploiting model internals}. Specifically, we expect that leveraging the model’s own gradients during training to perturb inputs across multiple percentages and penalizing instability in confidence‑drop trajectories will suppress the attack feature. We further hypothesize that \textit{enforcing distributional self‑consistency through a KL‑based regularization term will prevent distributional shifts that occur when inputs are perturbed}.
Motivated by these hypotheses, we propose a defense against attribution‑guided perturbation membership inference attacks, which we call \textbf{\textit{Trajectory-Invariant Explanation Regularization (TIER)}}.

To formalize our intuition, we pose the following research questions: 
\begin{itemize}
 \item RQ1: To what extent does the coupling of trajectory invariance and distributional self-consistency suppress membership-dependent leakage in attribution profiles?
 \item RQ2: How does the proposed regularization mitigate sensitivity to input perturbations and prevent distributional shifts?
 \item RQ3: Can the TIER framework achieve a Pareto-optimal balance between membership privacy and model utility?
 \item RQ4: Does the defense preserve the semantic interpretability and functional sufficiency of model explanations?
 \item RQ5: Is the TIER defense resilient to architectural variations and hyperparameter sensitivity?
 
\end{itemize}

\textbf{Contributions:} The principal contributions of this work are as follows:
\begin{enumerate}

\item[{1.}] We study membership-dependent explanation sensitivity— \textit{the phenomenon that member and non-member samples exhibit different prediction trajectories under explanation-guided perturbations} —and propose a training-time regularizer, TIER, that suppresses the separability of these trajectories.

\item[{2.}] We conduct extensive experiments to quantify the privacy risks of deep learning models, and highlight the privacy–utility trade‑offs using attack success rates, utility measures, and explanation fidelity metrics.

			\item[{3.}] We demonstrate that our defense generalizes across five benchmark attack datasets, two target model architectures, and seven state‑of‑the‑art attribution‑based explanation methods, successfully lowering the attack success while preserving both model utility and explanation fidelity to a substantial degree.

\end{enumerate}

\begin{figure}[htbp]
    \centering
    \includegraphics[width=\columnwidth]{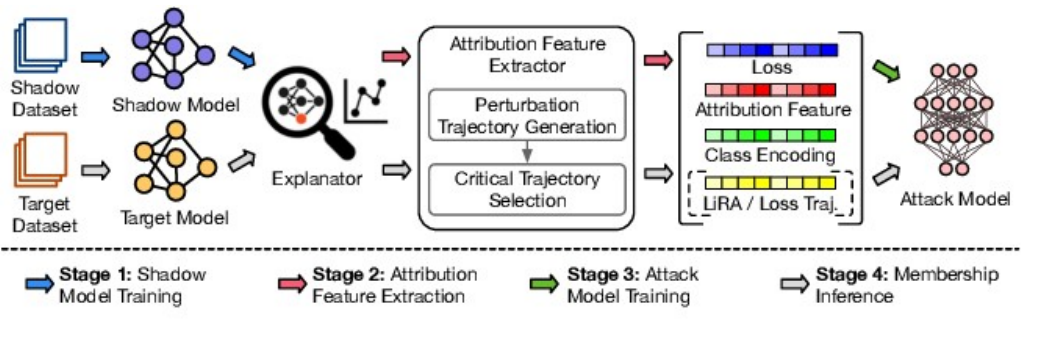}
    \caption{Overview of the threat model and attack pipeline, illustrating the four stages from shadow model training to membership inference (reproduced from \cite{10646875}).}
    \label{fig:thm}    
\end{figure}

\section{Related Work}
\label{sec:relatedwork}
\subsection{Membership Inference Attacks}
While membership inference has been extensively studied across deep learning\cite{9793586}, GANs\cite{10221704}, GNNs\cite{11072027}, and time-series models\cite{11269854}, the privacy risks of explainable ML have received comparatively less attention. Shokri et al.\ were the first to demonstrate that variance in backpropagation-based explanations can reveal membership status \cite{shokri2021privacyrisksmodelexplanations}, while Pawelczyk et al. designed attacks on counterfactual explanations in binary classifiers \cite{pawelczyk2021exploringcounterfactualexplanationslens}. Other works extend to feature inference \cite{Luo_2022}, attribute inference \cite{duddu2022inferringsensitiveattributesmodel}, and model inversion attacks enhanced by explanations \cite{zhao2022exploitingexplanationsmodelinversion}. More recently, \cite{10646875} developed a shadow-model attack that leverages perturbations guided by feature attribution explanations, systematically demonstrating strong success at identifying specific training data members in deep learning settings. This attack represents the state of the art in explainability-driven membership inference and forms the basis of the threat model we defend against.
\subsection{Membership inference defenses:}
Given the serious privacy risks posed by membership inference attacks, several defense strategies have been proposed at both the model and output levels. Differential Privacy (DP)\cite{Abadi_2016} is the most widely adopted mechanism, with DP-SGD introducing noise during training to constrain the ability to distinguish between datasets differing by a single record, thereby mitigating membership leakage. In contrast, MemGuard \cite{10.1145/3319535.3363201} operates at inference time by perturbing confidence scores, converting them into adversarial examples that confound attack models and disrupt reliable membership inference. Beyond these baselines, recent works have explored knowledge distillation approaches such as C2DP \cite{Yang2025C2DPCK}, which leverages CLIP-conditioned distillation to improve utility while reducing reliance on private training data, and PFE-KD \cite{articleLiu}, which integrates feature extraction with distillation to reduce attack accuracy without sacrificing model performance. Other defenses include metric mapping \cite{Yadav2025MitigatingBM}, which reconstructs prediction vectors to align member and non-member distributions, and Ensemble Partitioning Defense (EPD)\cite{10.1007/978-981-96-9872-1_36}, which partitions training data across sub-models and standardizes confidence scores during inference.

Despite these advances, existing defenses remain inadeguate against attribution-guided perturbation attacks. Differential privacy  often degrades model utility, while inference-time defenses like Memguard are ineffective against attacks that exploit attribution maps. Knowledge distillation and ensemble methods improve utility but do not directly address attribution-driven leakage. In contrast, our work explicitly targets attribution-guided perturbation-based membership inference.

\section{Preliminaries and Threat Model}
\label{sec:prelims}
This section establishes the formal notation and the adversarial framework used in our study. Our threat model and perturbation strategies follow established membership inference protocols in \cite{10646875} and are illustrated in Figure~\ref{fig:thm}.

\section{Threat Model}
\label{sec:threat_model}

\subsection{Overview and Scenario}
To contextualize the risk, we consider a scenario where a machine learning model is deployed as a service (MLaaS), providing both predictions and explanations. For instance, a medical diagnostic tool might provide a classification along with an attribution map highlighting the features (e.g., specific regions of an X-ray) that influenced the decision. While the model's internal parameters and training data are private, an adversary seeks to determine if a specific individual’s data was used to train the model. By observing how these explanations change as the input is slightly modified, an attacker can identify ``memorization signatures'' unique to the training set, thereby compromising the individual's membership privacy.

\subsection{Problem Formulation and Notation}
Following the framework of \cite{10646875}, we formalize this as a black-box membership inference attack. Let $\mathcal{D}_{tr}^{tg}$ denote the private target training set. We define the input space as $\mathcal{X}$ and the label space as $\mathcal{Y}=\{1,\dots,K\}$. A target classifier $f(x; \theta): \mathcal{X} \to \mathbb{R}^K$, parameterized by $\theta$, outputs a vector of confidence scores $P(y|x)$ for each class $y$. 

\paragraph{Adversary's Knowledge and Shadow Training:} 
The adversary lacks access to the target model parameters $\theta$ or the private dataset $\mathcal{D}_{tr}^{tg}$. However, they can query $f$ to obtain confidence scores $P(y|x)$ and attribution maps $A(x,f)$. To overcome the lack of direct access, the adversary employs a \textit{shadow model} approach. 

The adversary is assumed to possess auxiliary data $\mathcal{D}_{tr}^{sh}$, referred to as a \textbf{shadow dataset}, drawn from the same distribution as the target data. The adversary uses this to train a \textbf{shadow model} $f_S$, which serves as a proxy to simulate the target model's behavior. By performing attacks on $f_S$—where the ground-truth membership is known—the adversary learns to distinguish between members $(x,y) \in \mathcal{D}_{tr}^{sh}$ and non-members $(x,y) \in \mathcal{D}_{te}^{sh}$. Throughout this paper, superscripts $^{tg}$, $^{sh}$, and $^{at}$ refer to the target, shadow, and attack models, while subscripts $_{tr}$ and $_{te}$ denote training and testing partitions, respectively.

\paragraph{Adversary's Strategy (Attribution-Guided Perturbations):}
The adversary exploits $A(x,f)$ to systematically perturb an input $x$. To prioritize features, the attacker computes an ordering:
\begin{equation}
    I_{ij} = g_{ij} - \alpha \cdot TV(m_{ij})
\end{equation}
where $g_{ij}$ is the attribution for the position $(i,j)$ in the original image $m_{ij}$, and $TV(m_{ij})$ denotes the local total variation. Using either \textit{Most Relevant First} (MoRF) or \textit{Least Relevant First} (LeRF) strategies, the adversary iteratively applies a perturbation operator, yielding a \textit{confidence-drop trajectory}:
\begin{equation}
    \mathcal{T}(x) = \{P(y|x_0), P(y|x_1), \dots, P(y|x_n)\}
\end{equation}
where $x_0=x$ and $x_i$ is the input at perturbation step $i$.

\paragraph{Attack Execution:}
The adversary constructs a feature vector $\mathbf{v} = \mathcal{T}_{MoRF} \oplus \mathcal{T}_{LeRF} \oplus \mathcal{L}_{ce} \oplus O(c)$, where $\mathcal{L}_{ce}$ is the cross-entropy loss and $O(c)$ is the one-hot encoding of the predicted class. This vector is passed through an attack classifier $h_{atk}$ trained on shadow trajectories. Welch's $t$-test is applied to retain only trajectories with statistically significant ($\alpha=0.05$) discriminative power.
\begin{figure}[htbp]
    \centering
    \includegraphics[width=\columnwidth]{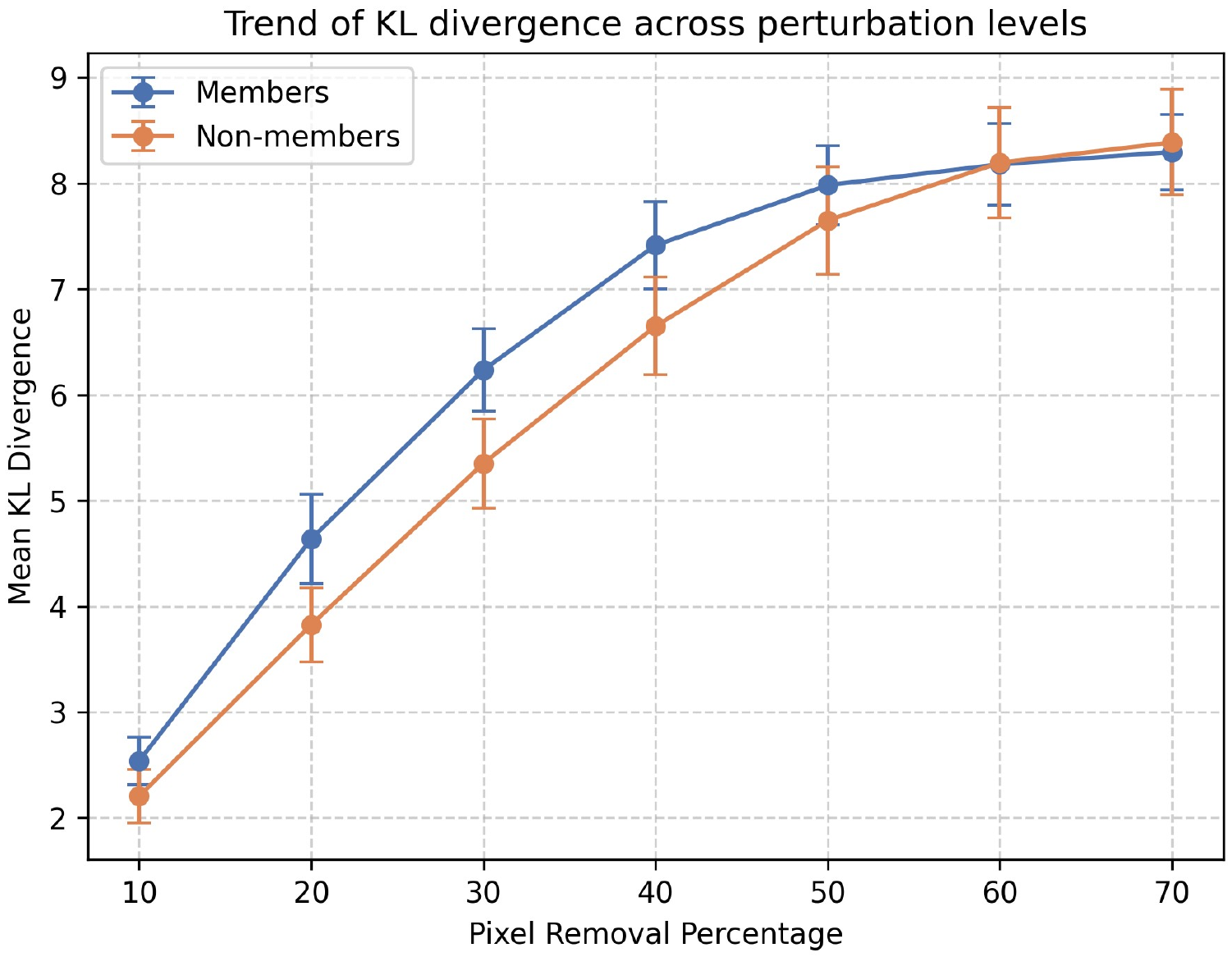}
    \caption{Trend of mean KL divergence for members and non-members across incremental pixel removal levels. Members exhibit a consistently higher divergence profile compared to non-members, particularly in the $10\%$--$40\%$ range, indicating a greater sensitivity to the removal of high-attribution features. Error bars represents the standard deviation across samples. \scriptsize(\textbf{Target Model: Resnet18, Dataset: CIFAR100})} 
    \label{fig:KL-div}    
\end{figure}

\section{Proposed Defense: Trajectory Invariant Explanation Regularization}
\label{sec:proposed}

\subsection{Motivation: Distributional Divergence in Membership Inference}
\label{subsec:mot}
Our defense is motivated by the observation that membership status alters model sensitivity to systematic perturbations. To quantify this sensitivity gap, we define an Attribution-Guided Perturbation (AGP) protocol:

\begin{enumerate}
\item \textbf{Attribution Mapping:} For input $x$, compute gradients of the target logit w.r.t. pixels. The attribution map $M$ is the absolute mean gradient across channels, highlighting influential pixels.
\item \textbf{Mean-Filling Perturbations:} Generate perturbed inputs $\{\tilde{x}_k\}$ by masking the top $k\%$ pixels ($k \in \{10,20,\dots,70\}$) ranked by $M$, replacing them with the image’s mean intensity.
\item \textbf{Divergence Profiling:} Measure the KL divergence between the clean prediction $P=f(x)$ and each perturbed prediction $Q_k=f(\tilde{x}_k)$:

\[
D_{KL}(P \parallel Q_k) = \sum_i P(i)\log\frac{P(i)}{Q_k(i)}.
\]

\end{enumerate}
As illustrated in Figure~\ref{fig:KL-div}, the mean KL divergence increases across all samples as more information is removed. However, a significant gap exists in the divergence profiles: members (\textit{samples from the training set}) exhibit a sharper increase in divergence compared to non-members (\textit{samples from the test set}), particularly at lower removal percentages ($10\%$--$40\%$). This suggests that models rely on brittle, memorized features for training samples which, when even slightly perturbed, cause a rapid collapse in confidence. To statistically validate this separation, we applied both the independent two‑sample $t$‑test and the Kolmogorov–Smirnov (KS) test at each removal level, confirming a robust distributional gap. As summarized in Table~\ref{tab:kl_stats_simple}, at $10\%$ pixel removal, members show a significantly higher mean KL divergence ($\textbf{2.6}$) than non‑members ($\textbf{2.2}$), supported by a $t$‑statistic of $\textbf{7.9}$ ($p \approx \textbf{1.1} \times \textbf{10}^\textbf{-12}$) and a KS statistic of $D = \textbf{0.5}$ ($p \approx \textbf{5.7} \times \textbf{10}^\textbf{-13}$). Notably, this gap peaks at $20\%$ removal ($t = \textbf{14.3}, D = \textbf{0.8}$) before the profiles converge at extreme removal levels (above $50\%$), where predictive capacity is exhausted for both groups.

\begin{table}[t]
\centering
\footnotesize 
\setlength{\tabcolsep}{4pt}
\begin{tabular}{@{}lccc@{}}
\toprule
\textbf{\% Rem.} & \textbf{Mean (M, NM)} & \textbf{t-test ($t, p$)} & \textbf{KS-test ($D, p$)} \\ \midrule
10\% & 2.6, 2.2 & $7.9, 1.1 \times 10^{-12}$ & $0.5, 5.7 \times 10^{-13}$ \\
20\% & 4.6, 3.8 & $14.3, 3.2 \times 10^{-30}$ & $0.8, 6.4 \times 10^{-26}$ \\
30\% & 6.2, 5.4 & $13.3, 8.8 \times 10^{-28}$ & $0.7, 2.5 \times 10^{-22}$ \\
40\% & 7.4, 6.7 & $10.9, 2.7 \times 10^{-21}$ & $0.7, 1.8 \times 10^{-20}$ \\
50\% & 8.0, 7.7 & $4.9, 2.9 \times 10^{-6}$ & $0.3, 9.2 \times 10^{-5}$ \\ \bottomrule
\end{tabular}
\caption{KL divergence comparison between members (M) and non-members (NM). Statistical significance confirms distinct distribution shifts.}
\label{tab:kl_stats_simple}
\end{table}

\subsection{Research Hypothesis}

The divergence gap identified in Section~\ref{subsec:mot} motivates the following hypothesis:

\begin{quote}
    \textbf{Hypothesis:} \textit{If instability and distributional shifts in confidence trajectories are the primary discriminative cues exploited by a membership inference adversary, then suppressing these signals will align member behavior with that of non-members, thereby mitigating privacy leakage.}
\end{quote}

Our rationale is threefold. First, \textit{the adversary exploits instability in confidence trajectories}; therefore, \textit{stability must be enforced.} Second, \textit{the adversary relies on distributional shifts;} therefore, \textit{distributional consistency must be promoted.} Third, the natural way to achieve both is to \textit{simulate attribution-guided perturbations} internally during training itself, using gradient-based saliency maps, and \textit{penalize the same signals the attacker would exploit}. 

\subsection{Design Goals}

To translate our hypothesis into a functional defense, we identify three key requirements:

\begin{enumerate}
    \item \textbf{Privacy Robustness:} The defense should significantly reduce the success rate of membership inference attacks by narrowing the gap between member and non-member distributions.
    
    \item \textbf{Utility Preservation:} The defense should limit accuracy degradation on clean data and preserve explanation fidelity, ensuring privacy gains without undermining the model’s practical utility or explanation usefulness.
    
    \item \textbf{Seamless Integration:} The defense should provide a plug-and-play solution that integrates directly into standard supervised training pipelines without requiring architectural modifications.
\end{enumerate}

\subsection{Formal Definition}
\label{subsec:defn}
We propose a specialised \textbf{Trajectory-Invariant Explanation Regularization (TIER)} defense that treats the model’s own gradients as a primary signal during training rather than a post-hoc analysis tool and enforces a predictable and stable decay in model confidence when pixels are removed in order of their calculated importance.

Our proposed training procedure is detailed in Algorithm~\ref{alg:tier_training}. To the standard supervised cross-entropy loss $\mathcal{L}_{ce}$, we add a regularization term weighted by the coefficient $\lambda$. This regularization consists of two primary components:

\begin{itemize}
    \item \textbf{Distributional Stability Term} $\mathcal{L}_{kl}$, which enforces self-consistency between clean and perturbed predictions via Kullback-Leibler divergence.
    \item \textbf{Trajectory Stability Term} $\mathcal{L}_{var}$, which minimizes the variance of confidence-drop trajectories across perturbation levels.
\end{itemize}

Formally, the total objective function is defined as:

\begin{equation}
\label{eq:total_loss}
\mathcal{L}_{total} = \mathcal{L}_{ce} + \lambda \cdot \big( \mathcal{L}_{kl} + \beta \cdot \mathcal{L}_{var} \big).
\end{equation}

Here, $\mathcal{L}_{ce}$ denotes the supervised cross-entropy loss, $\mathcal{L}_{\text{KL}}$ penalizes distributional shifts between clean and perturbed predictions, $\mathcal{L}_{\text{var}}$ penalizes instability in confidence-drop trajectories, and $\beta$ hyperparameter balances their contributions. Together, these terms aim to align the explanation profiles of members and non-members.

\begin{algorithm}[htbp]  
\footnotesize
\SetAlgoLined

\KwIn{Model $\theta$, Training Data $\mathcal{D}$, Levels $P=\{5, 10, 20\}$, Hyperparameters $\beta, \lambda_e$, Augmentation Pipeline $\mathcal{T}$}
\KwOut{Optimized Model $\theta$}

\For{epoch $e = 1$ \KwTo $E$}{
    \For{batch $(x, y) \in \mathcal{D}$}{
        \textbf{Apply Data Augmentation:} $x' = \mathcal{T}(x)$\;
        Compute Attribution on augmented input: $A = |\nabla_{x'} f_y(x')|$\;
        
        \If{$e \geq \text{warmup}$}{
            \For{each level $p \in P$}{
                Generate $\tilde{x}'_p$ by masking top-$p\%$ features in $A$\;
                
                Compute KL divergence:\\
                $\mathcal{L}_{KL}^{(p)} = \text{KL}(\text{Softmax}(f(x')) \parallel \text{Softmax}(f(\tilde{x}'_p)))$\;
                
                Compute the drop in confidence:\\
                $\delta_p = f_y(x') - f_y(\tilde{x}'_p)$\;
            }
            
            $\mathcal{L}_{KL} = \text{mean}_{p \in P}(\mathcal{L}_{KL}^{(p)})$\;
            
            Compute the average variance of confidence drops:\\
            $\mathcal{L}_{var} = \text{mean}_{p \in P} (\text{Var}_{batch}(\delta_p))$\;
            
            Total loss with TIER constraints:\\
            $\mathcal{L}_{total} = \mathcal{L}_{CE}(f(x'), y) + \lambda_e (\mathcal{L}_{KL} + \beta \mathcal{L}_{var})$\;
        }
        \Else{
            Standard training with augmented data:\\
            $\mathcal{L}_{total} = \mathcal{L}_{CE}(f(x'), y)$\;
        }
        Update $\theta$ via Backpropagation using $\nabla_{\theta} \mathcal{L}_{total}$\;
    }
}
\caption{Trajectory-Invariant Explanation Regularization.}
\label{alg:tier_training}
\end{algorithm}

\subsection{Defense Components}
\label{subsec:defense_components}
The TIER framework is composed of two synergistic components designed to mask membership-specific information in explanation manifolds: the \textit{Intrinsic Regularization} of trajectories and \textit{Extrinsic Semantic Augmentation}.

\subsubsection{\textbf{Trajectory-Invariant Regularization}}
The core of our defense lies in the simultaneous optimization of distributional self-consistency ($\mathcal{L}_{kl}$) and trajectory stability ($\mathcal{L}_{var}$). [Insert brief math/logic here]. This stage is responsible for the internal smoothing of the model manifold.

\subsubsection{\textbf{Semantic Data Augmentation}}
As a complementary stage, we introduce a stochastic augmentation pipeline $\mathcal{T}$. While the regularization terms smooth the model's internal response, the objective of this augmentation stage is to ensure that attribution maps $A$ remain invariant to spatial perturbations. By applying random horizontal flipping, rotations, and cropping, we prevent the importance scores from overfitting to specific coordinates, thereby decoupling the explanation from membership-dependent spatial signatures.

As discussed in the Ablation Study (Section~\ref{subsec:ablation}), this configuration allows us to isolate the specific contributions of the regularization-only variant versus the full defense.

\subsection{Implementation Details}
\label{subsec:implementation}
As detailed in Algorithm~\ref{alg:tier_training}, we introduce a short warm-up phase during the initial 5 epochs. This allows the classifier to capture primary semantic features and stabilize its decision boundary before the $\mathcal{L}_{kl}$ and $\mathcal{L}_{var}$ constraints are enforced. We use a batch size of 128 and train for 100 epochs, maintaining consistency with the protocol in \cite{10646875}. The stability terms are computed over a discrete set of perturbation levels $P=\{5, 10, 20\}$, representing the percentage of top-attribution pixels removed. 

\section{Experimental Setup}
\label{sec:exp}
We formalize the experimental framework by defining adversarial and defensive roles, followed by model, dataset, and metric specifications.

\subsection{Conceptual Framework and Model Roles}
To evaluate membership privacy rigorously, we categorize our setup into two distinct roles: the Attacker and the Defender. Each party is responsible for building a specific suite of models to fulfill their objectives.

\subsubsection{Attacker's Role: Membership Inference}
The Attacker operates in a black-box setting with no access to the Target Model’s training data or internal parameters. To perform the attack, the Attacker constructs:
\begin{itemize}
    \item \textbf{Shadow Model ($f_S$):} Trained on a proxy dataset to mimic the Target Model's behavior. It serves as a ground-truth generator where "member" and "non-member" labels are known to the Attacker.
    \item \textbf{Attack Model ($\mathcal{A}$):} A binary classifier trained on features as described in Section~\ref{sec:threat_model} extracted from the Shadow Model. Its goal is to learn the statistical signatures that distinguish seen data from unseen data.
\end{itemize}

\subsubsection{Defender's Role: Privacy Preservation}
The Defender manages the lifecycle of the primary application model. The Defender constructs:
\begin{itemize}
    \item \textbf{Target Model ($f_T$):} The model providing the end-user utility (e.g., image classification). We evaluate $f_T$ in two states: a \textit{baseline} (no defense) and a \textit{defended} version utilizing our proposed trajectory-invariant regularization.
    \item \textbf{Internal Audit Classifier:} To validate the defense during development, the Defender builds an internal membership inference classifier (following the Attacker’s methodology) to measure the reduction in successful inferences.
\end{itemize}

\subsection{Datasets and Disjoint Splits}

\subsubsection{Datasets}
We evaluate our defense across five diverse image datasets, categorized into established benchmarks and specialized real-world scenarios. For baseline comparison with prior membership inference research, we utilize four widely adopted benchmarks: CIFAR-10~\cite{CIFAR10}, CIFAR-100~\cite{CIFAR10}, CINIC-10~\cite{darlow2018cinic10imagenetcifar10}, and the German Traffic Sign Recognition Benchmark (GTSRB)~\cite{Stallkamp-IJCNN-2011}. 

To extend our evaluation beyond standard classification tasks, we additionally include the OpenForensics dataset~\cite{ltnghia-ICCV2021}, a large-scale benchmark specifically designed for multi-face forgery detection. The inclusion of OpenForensics allows us to assess privacy risks in high-stakes, contemporary domains such as deepfake detection and facial forensics.

\subsubsection{Splits and Attacker Dataset Construction}
To prevent data leakage and maintain comparability with prior work, we follow the dataset construction protocol of established attack studies~\cite{10646875, shokri2021privacyrisksmodelexplanations}. We construct strictly disjoint splits by first partitioning the sampled source data into two primary halves: one reserved for the Defender (Target) and one for the Attacker (Shadow).

Each half is further divided into training (Members) and test (Non-members) portions, yielding four mutually exclusive subsets: $\mathcal{D}_{tr}^{tg}$, $\mathcal{D}_{te}^{tg}$, $\mathcal{D}_{tr}^{sh}$, and $\mathcal{D}_{te}^{sh}$. The training sets are used to fit the target and shadow models, respectively, while the corresponding test sets provide the attacker with balanced access to non-member samples. For instance, in CIFAR-100, we sample 40,000 images and split them evenly into 10,000-sample portions for each of the four disjoint sets, as detailed in Table~\ref{tab:data_splits}.

\begin{table}[ht]
\centering
\footnotesize 
\setlength{\tabcolsep}{4.5pt} 
\begin{tabular}{lcccccc}
\toprule
\textbf{Dataset} & \textbf{Target} & \textbf{Test} & \textbf{Shadow} & \textbf{S-Test} & \textbf{Attack} & \textbf{A-Test} \\
& \scriptsize ($ \mathcal{D}_{tr}^{tg} $) & \scriptsize ($ \mathcal{D}_{te}^{tg} $) & \scriptsize ($ \mathcal{D}_{tr}^{sh} $) & \scriptsize ($ \mathcal{D}_{te}^{sh} $) & \scriptsize ($ \mathcal{D}_{tr}^{at} $) & \scriptsize ($ \mathcal{D}_{te}^{at} $) \\
\midrule
CIFAR-10    & 10,000 & 10,000 & 10,000 & 10,000 & 16,000 & 4,000 \\
CIFAR-100   & 10,000 & 10,000 & 10,000 & 10,000 & 16,000 & 4,000 \\
CINIC-10    & 10,000 & 10,000 & 10,000 & 10,000 & 16,000 & 4,000 \\
GTSRB       & 1,500  & 1,500  & 1,500  & 1,500  & 2,400  & 600   \\
OForensics  & 5,000  & 5,000  & 5,000  & 5,000  & 8,000  & 2,000 \\
\bottomrule
\end{tabular}
\caption{Data splits across evaluated datasets. Each set is partitioned to ensure disjoint distributions for target, shadow, and attack model training/evaluation.}
\label{tab:data_splits}
\end{table}

\subsection{Experimental Flow}
\label{subsec:flow}
\paragraph{Attacker's Flow (Threat Model).} The Attacker’s workflow follows a systematic three-stage process to extract and exploit membership signatures:

\begin{enumerate}
    \item \textbf{Shadow Environment Construction:} The Attacker trains the Shadow Model $f_S$ on $\mathcal{D}_{tr}^{sh}$. This model acts as a proxy to observe how a specific architecture reacts to ``memorized'' (member) versus unseen (non-member) data.
    \item \textbf{Attack Training and Feature Extraction:} The Attacker labels $\mathcal{D}_{tr}^{sh}$ as members ($1$) and $\mathcal{D}_{te}^{sh}$ as non-members ($0$). These subsets are combined and randomly split into attacker training ($\mathcal{D}_{tr}^{at}$) and evaluation ($\mathcal{D}_{te}^{at}$) sets. For larger datasets (e.g., CIFAR-10/100, CINIC), this corresponds to roughly 16k/4k samples (Table~\ref{tab:data_splits}). This construction guarantees that the Attacker's model relies solely on shadow model behavior, without any overlap with the target data.
    For each sample, attribution maps are generated (e.g., via Grad-CAM or SHAP) and MoRF/LeRF trajectories are extracted to form a tabular dataset, as detailed in Section~\ref{sec:threat_model}. The Attack Model $\mathcal{A}$ is then trained on these features to learn the statistical decision boundary of membership.
    \item \textbf{Inference Deployment:} Finally, $\mathcal{A}$ is deployed in a black-box manner against the Target Model $f_T$. The Attacker feeds target samples into $f_T$, extracts the corresponding attribution trajectories, and utilizes $\mathcal{A}$ to predict the membership status.
\end{enumerate}
\begin{figure*}[htbp]
\centering
\begin{subfigure}[b]{0.48\textwidth}
    \includegraphics[width=\textwidth]{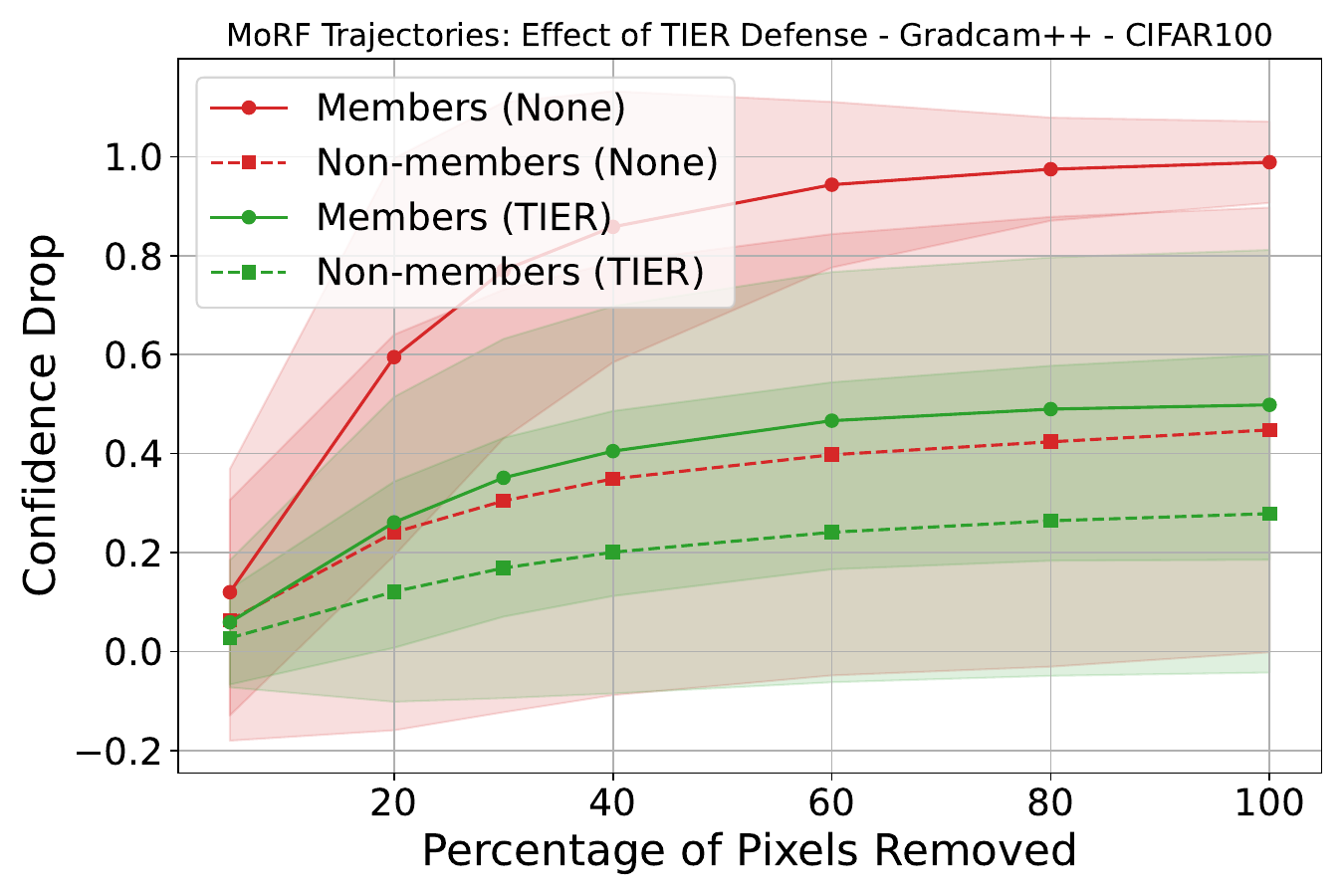}
\end{subfigure}
\begin{subfigure}[b]{0.48\textwidth}
    \includegraphics[width=\textwidth]{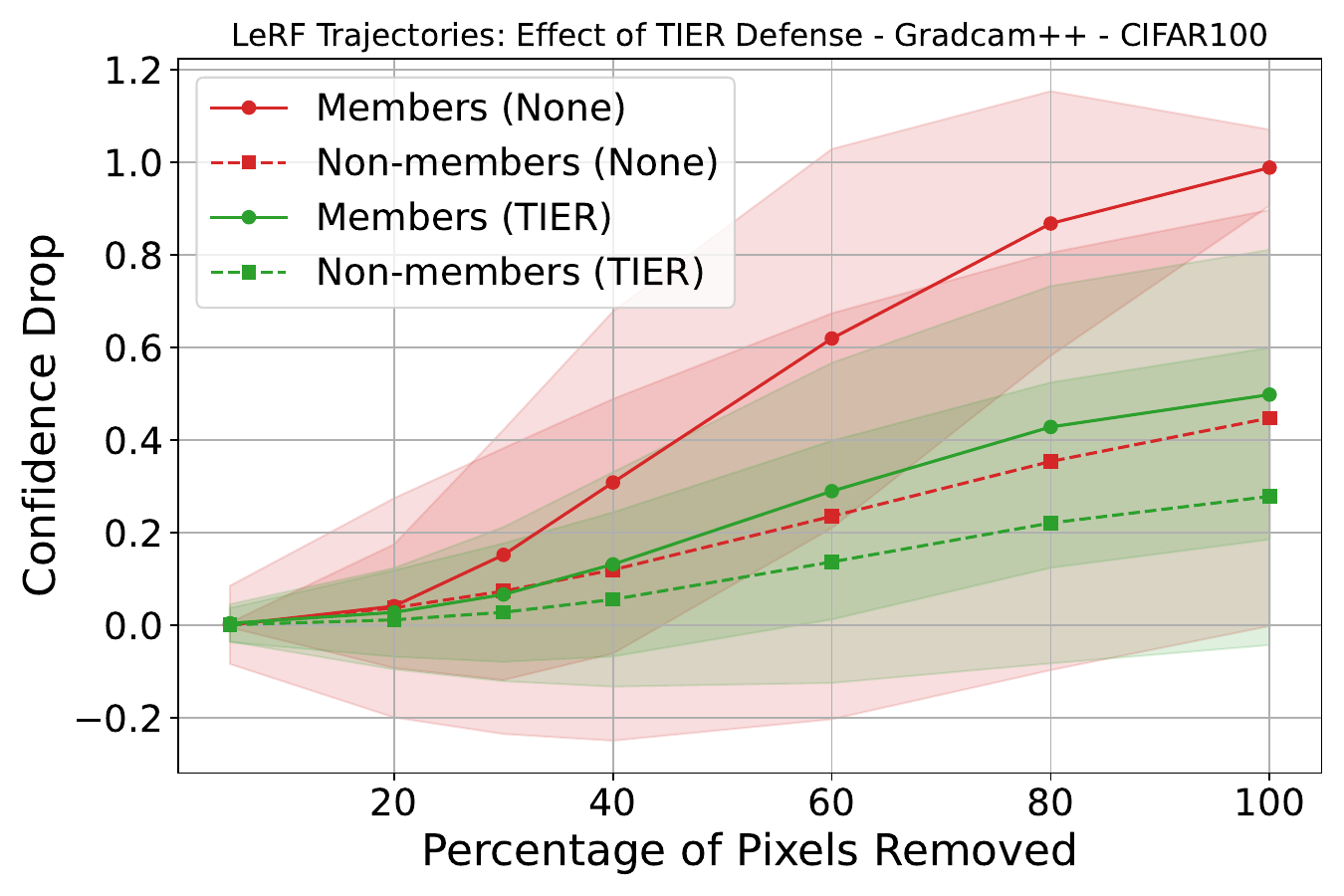}
\end{subfigure}
\begin{subfigure}[b]{0.48\textwidth}
    \includegraphics[width=\textwidth]{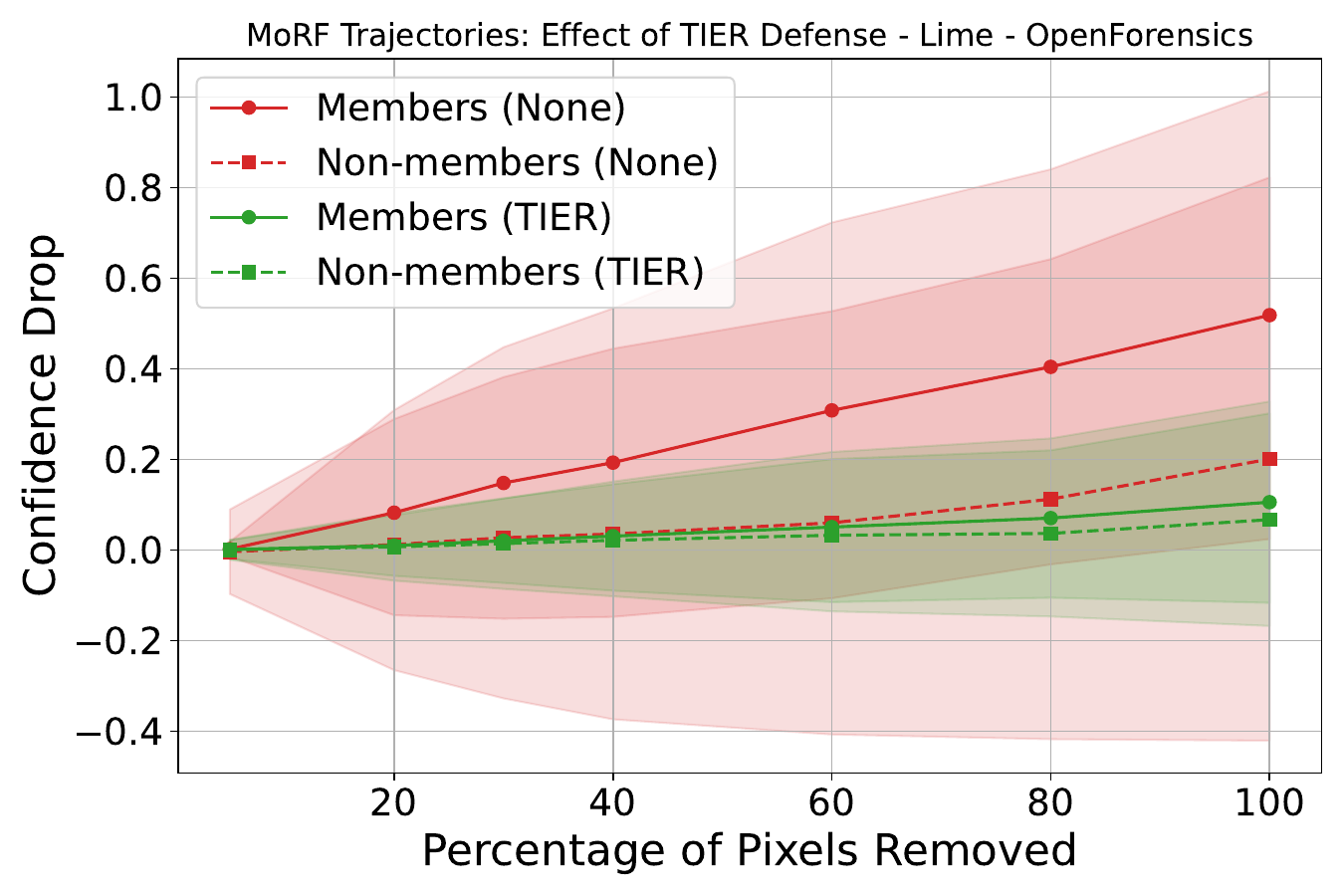}
\end{subfigure}
\begin{subfigure}[b]{0.48\textwidth}
    \includegraphics[width=\textwidth]{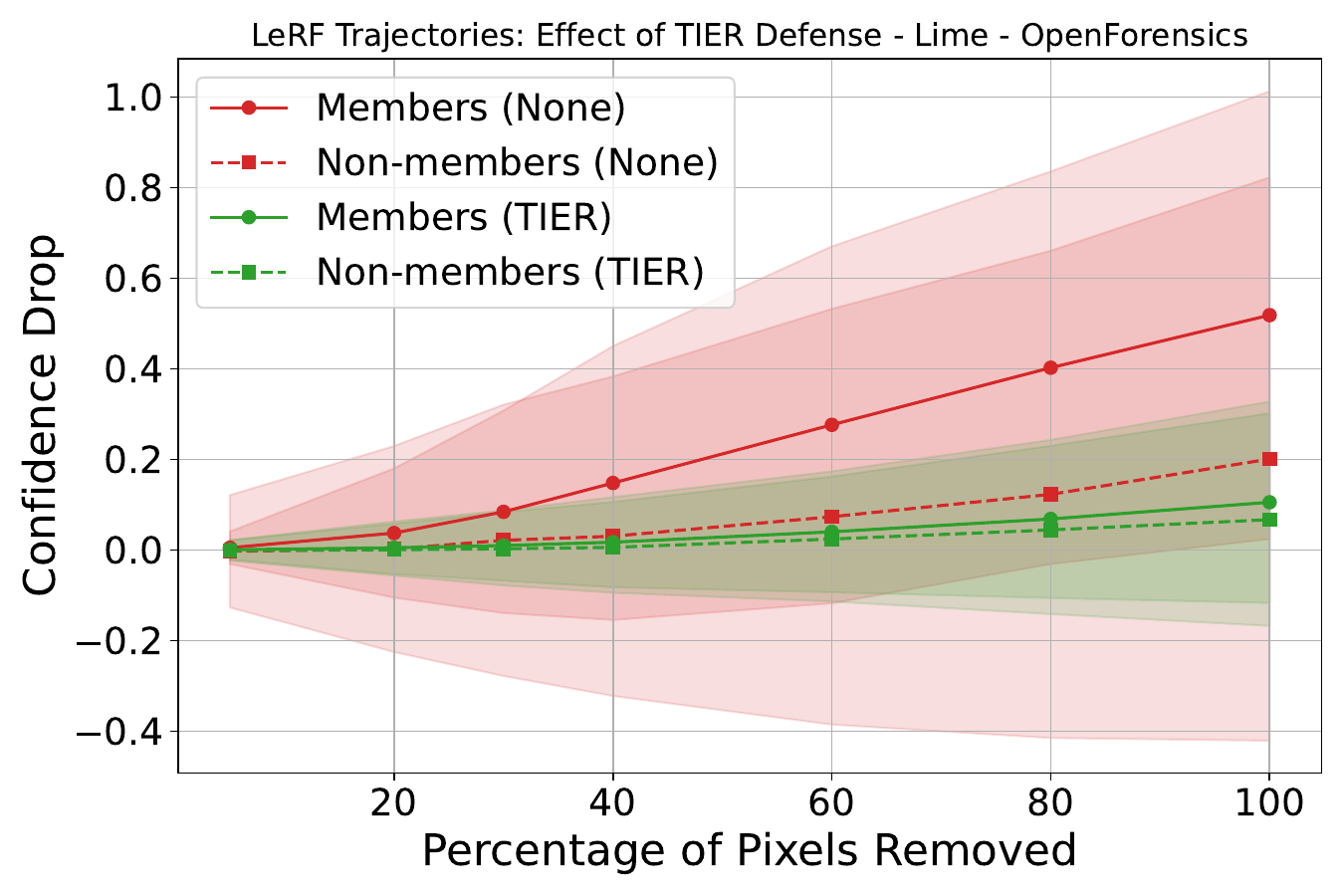}
\end{subfigure}
\caption{Evaluation of TIER defense via MoRF and LeRF confidence-drop trajectories on CIFAR-100 (top) and OpenForensics (bottom). The minimized gap between Member (solid) and Non-member (dashed) curves under the TIER defense (green), compared to the distinct separation in the undefended baseline (red), demonstrates the effective masking of membership-specific statistical signatures.}
\label{fig:main_traj}
\end{figure*}

\paragraph{Defender's Flow.} The Defender's workflow follows a three-stage pipeline to balance utility and privacy:

\begin{enumerate}
    \item \textbf{Baseline Establishment:} The Defender first trains the target model $f_T$ on $\mathcal{D}_{tr}^{tg}$ without defense to establish baseline utility and privacy leakage metrics.
    \item \textbf{Defense Integration:} The Trajectory-Invariant Explanation Regularization (TIER) is integrated into the $f_T$ training loop. This stabilizes the confidence manifold, ensuring attribution trajectories for members ($\mathcal{D}_{tr}^{tg}$) and non-members ($\mathcal{D}_{te}^{tg}$) remain statistically indistinguishable.
    \item \textbf{Internal Privacy Audit:} To validate efficacy, the Defender mimics the adversary by training an internal membership classifier. Following the established protocol, $\mathcal{D}_{tr}^{tg}$ and $\mathcal{D}_{te}^{tg}$ are combined and partitioned into internal training ($80\%$) and evaluation ($20\%$) sets. A successful defense is confirmed by a significant reduction in internal classifier accuracy compared to the baseline, signifying mitigated membership leakage.
\end{enumerate}

\subsection{Model Architectures}
For our quantitative analysis, we primarily adopt ResNet-18~\cite{he2015deepresiduallearningimage} as the backbone for both target and shadow models, representing a standard high-performance architecture utilizing residual connections. To evaluate the architectural generalizability and robustness of our proposed defense across varying model complexities, we further include MobileNetV2~\cite{sandler2019mobilenetv2invertedresidualslinear}. 

MobileNetV2 was specifically selected to assess the efficacy of the defense on resource-constrained architectures characterized by inverted residuals and linear bottlenecks. By testing both a standard deep residual network and a mobile-centric efficiency-optimized model, we demonstrate that the proposed defense remains effective regardless of parameter density or architectural design. All models are trained using standard regularization strategies, including weight decay~\cite{NIPS1991_8eefcfdf} and data augmentation~\cite{cubuk2019autoaugmentlearningaugmentationpolicies}.

\subsection{Evaluation Metrics}
We evaluate the privacy-utility-interpretability tradeoff across three dimensions:

\noindent\textbf{Attack Performance (Privacy):} 
To quantify the success of the inference attacker $\mathcal{A}$ in distinguishing members from non-members, we report:
\begin{itemize}
    \item \textbf{TPR at low FPR (0.1\%):} Measures the attacker's ability to identify members with high confidence ($\downarrow$).
    \item \textbf{Balanced Accuracy and AUC:} Assess the overall binary classification performance of $\mathcal{A}$ across the evaluation set ($\downarrow$).
\end{itemize}

\noindent\textbf{Explanation Quality (Interpretability):}
To ensure the defense preserves attribution reliability, we evaluate two distinct properties:
\begin{itemize}
    \item \textbf{Axiomatic Quality (Quantus Framework):} Using the \textbf{Quantus framework}~\cite{hedstrom2023quantus}, we measure the technical reliability of explanations via \textit{Faithfulness Correlation/Estimate} ($\uparrow$) and \textit{Sufficiency} ($\downarrow$). These are calculated independently for each model to quantify how strictly attribution maps correspond to internal logit shifts.
    \item \textbf{Visual and Structural Fidelity:} We utilize \textit{SSIM} ($\uparrow$), \textit{MSE} ($\downarrow$), and \textit{Cosine Similarity} ($\uparrow$). These are calculated by comparing the defended model's maps directly against the undefended baseline to quantify ``semantic intent'' preservation for human observers.
\end{itemize}

\noindent\textbf{Model Utility:} 
To monitor the impact on the primary task and efficiency, we record:
\begin{itemize}
    \item \textbf{Top-1 and Top-5 Accuracy ($\uparrow$):} Top-1 assesses strict classification, while Top-5 evaluates whether the defense prevents ``semantic collapse,'' ensuring the correct class remains within the highest confidence set.
    \item \textbf{Test Loss ($\downarrow$):} Provides a granular view of model confidence and generalization beyond binary thresholds.
    \item \textbf{Average Epoch Time:} Quantifies the computational overhead introduced by TIER regularization.
\end{itemize}

\section{Results and Discussions}
\subsection{Defense Mechanisms: Trajectory and Distributional Stability}
\label{subsec:results1}
\textbf{RQ1: To what extent does the coupling of trajectory invariance and distributional self-consistency suppress membership-dependent leakage in attribution profiles?} To evaluate defense efficacy, we analyze confidence-drop trajectories across five datasets and seven explanation methods. To provide a rigorous yet concise evaluation, we present representative trajectory plots for two distinct paradigms: gradient-based attribution (\textit{Grad-CAM++ on CIFAR-100}) and model-agnostic perturbation (\textit{Lime on OpenForensics}). 

As shown in Figure~\ref{fig:main_traj}, TIER achieves consistent gap compression across both MoRF and LeRF metrics, regardless of the underlying explanation logic. 
In the baseline model (red curves), a distinct gap between members (solid lines) and non-members (dashed lines) represents a primary leakage signal, where the model behaves differently for seen versus unseen data. Under the TIER defense (green curves), this gap is significantly minimized, and the trajectories become increasingly indistinguishable. This suggests that TIER effectively suppresses membership-specific statistical signatures, forcing the model's response to members to align closely with the non-member baseline.

Comprehensive plots for the remaining attribution methods and datasets are provided in the \textbf{\textit{Supplementary Material, Figure S2,S3,S4}}. Across all datasets, it can be seen that the coupling of trajectory stabilization and KL-regularized self-consistency effectively masks these membership-specific statistical signatures. Furthermore, by enforcing trajectory invariance, TIER smooths MoRF declines and forces member behavior to mirror the non-member baseline.

\begin{figure}[htbp]
    \centering
    \includegraphics[width=0.50\textwidth]{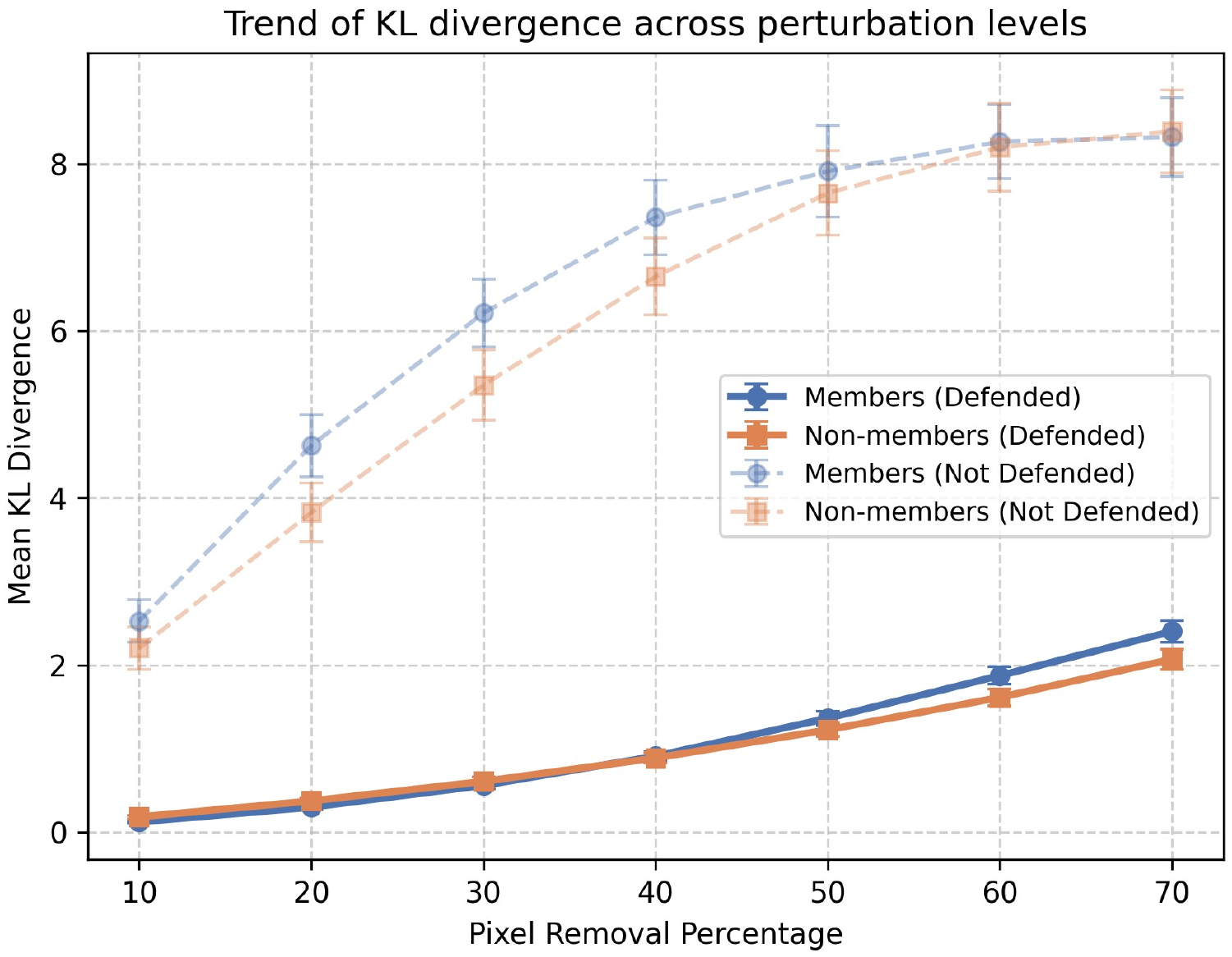}
    \caption{Trend of mean KL divergence for members and non-members after the application of the KL-regularization defense. The defense significantly suppresses the overall magnitude of divergence and introduces a stability crossover effect, where member samples exhibit improved resilience to low-level perturbations ($10\%$--$30\%$) compared to non-members. The convergence of profiles at moderate perturbation levels indicates a reduction in the statistical signatures typically exploited for membership inference. \scriptsize(\textbf{Target Model: Resnet18, Dataset: CIFAR100})} 
    \label{fig:KL-divdef}    
\end{figure}

\textbf{RQ2: How does the proposed regularization mitigate sensitivity to input perturbations and prevent distributional shifts?} TIER mitigates sensitivity to input perturbations by suppressing the high-magnitude divergence typically seen in overfit models. Furthermore, it prevents distributional shifts by aligning the attribution-guided trajectories of members and non-members, effectively neutralizing the unique statistical signatures used for inference. To evaluate this efficacy, we contrast the KL divergence profiles of the regularized model against the undefended baseline. As illustrated in Figure~\ref{fig:KL-divdef}, the defense achieves a \textit{multi-order suppression} of divergence magnitudes. While the undefended model exhibits a steep, divergent trajectory peaking at approximately $\textbf{8.4}$, the regularized model maintains a significantly compressed profile, peaking at only $\textbf{2.4}$ even under $\textbf{70}\%$ feature removal.

Beyond absolute magnitude, the most critical impact is the \textit{collapse of the membership gap}. In the baseline (dashed lines), members consistently exhibit a higher sensitivity to perturbations than non-members across all levels. TIER effectively eliminates this discriminative signature through a \textbf{stability crossover effect}: at lower perturbation levels ($10\%$--$30\%$), the membership-induced fragility is not only mitigated but inverted. For instance, at $10\%$ removal, members in the defended model exhibit a lower mean KL divergence ($\textbf{0.12}$) than non-members ($\textbf{0.18}$), reversing the baseline trend where members were significantly more volatile ($2.6$ vs $2.2$).

As perturbations reach the $40\%$ threshold, the profiles converge ($\textbf{0.92}$ vs. $\textbf{0.89}$), rendering the groups statistically indistinguishable. This convergence, validated by the t-test results in Table~\ref{tab:kl_stats_defense}, implies that TIER effectively "smoothes" the explanation manifold, depriving adversaries of the stable behavioral signatures required for high-confidence inference.

\begin{table}[t]
\centering
\footnotesize 
\setlength{\tabcolsep}{4pt}
\begin{tabular}{@{}lccc@{}}
\toprule
\textbf{\% Rem.} & \textbf{Mean (M, NM)} & \textbf{t-test ($t, p$)} & \textbf{KS-test ($D, p$)} \\ \midrule
10\% & 0.12, 0.18 & $-24.1, 1.3 \times 10^{-54}$ & $0.9, 2.3 \times 10^{-46}$ \\
20\% & 0.29, 0.37 & $-15.9, 5.4 \times 10^{-36}$ & $0.8, 5.1 \times 10^{-27}$ \\
30\% & 0.56, 0.60 & $-6.2, 3.6 \times 10^{-9}$ & $0.4, 1.5 \times 10^{-7}$ \\
40\% & 0.92, 0.89 & $3.1, 2.3 \times 10^{-3}$ & $0.3, 4.2 \times 10^{-3}$ \\
50\% & 1.36, 1.22 & $10.8, 7.1 \times 10^{-21}$ & $0.6, 9.8 \times 10^{-17}$ \\ \bottomrule
\end{tabular}
\caption{KL divergence comparison between members (M) and non-members (NM) across pixel removal percentages with \textbf{TIER} defense applied.}
\label{tab:kl_stats_defense}
\end{table}

\begin{table}[htbp]
\centering
\footnotesize 
\setlength{\tabcolsep}{3pt} 
\begin{tabular}{llcccc} 
\toprule
\textbf{Dataset} & \textbf{Defense} & \textbf{\shortstack{Train \\ Acc}} & \textbf{\shortstack{Target \\ (Top-1)}} & \textbf{\shortstack{Target \\ (Top-5)}} & \textbf{\shortstack{Mean \\ (Loss)}} \\ 
\midrule
\multirow{3}{*}{GTSRB} 
& None        & \textbf{1.00} & \textbf{0.75} & \textbf{0.94} & 0.97 \\
& DP ($\epsilon$=100) & 0.33 & 0.26 & 0.56 & 2.55 \\
& \textbf{TIER}        & 0.83 & \textbf{0.74} & \textbf{0.96} & \textbf{0.94} \\
\midrule
\multirow{3}{*}{CINIC-10} 
& None        & \textbf{1.00} & \textbf{0.63} & \textbf{0.96} & 2.32 \\
& DP ($\epsilon$=100) & 0.36 & 0.32 & 0.83 & 2.88 \\
& \textbf{TIER}        & 0.72 & \textbf{0.62} & \textbf{0.96} & \textbf{1.14} \\
\midrule
\multirow{3}{*}{CIFAR-10} 
& None        & \textbf{1.00} & \textbf{0.84} & \textbf{0.99} & 0.64 \\
& DP ($\epsilon$=100) & 0.41 & 0.40 & 0.86 & 1.80 \\
& \textbf{TIER}        & 0.82 & \textbf{0.77} & \textbf{0.98} & \textbf{0.67} \\
\midrule
\multirow{3}{*}{CIFAR-100} 
& None        & \textbf{1.00} & \textbf{0.48} & \textbf{0.77} & 3.53 \\
& DP ($\epsilon$=100) & 0.19 & 0.14 & 0.34 & 5.69 \\
& \textbf{TIER}        & 0.74 & \textbf{0.42} & \textbf{0.71} & \textbf{2.13} \\
\midrule
\multirow{3}{*}{OpenForensics} 
& None        & \textbf{0.99} & \textbf{0.70} & - & 1.59  \\
& DP ($\epsilon$=100) & 0.34 & 0.26 & - & 1.71 \\
& \textbf{TIER}        & 0.73 & \textbf{0.67} & - & \textbf{0.56} \\
\bottomrule
\end{tabular}
\caption{Utility metrics across different defenses. \textbf{Bold} values indicate performance within a competitive margin ($\pm 5\%$) of the baseline. Note the significant utility collapse for DP ($ \epsilon=100 $) compared to the stability of TIER. (\scriptsize \textit{OF has 2 classes; Top-5 not reported.})}
\label{tab:defense_utility}
\end{table}

\begin{table*}[t]
\centering
\footnotesize 
\setlength{\tabcolsep}{3.5pt} 
\begin{tabular}{@{}ll ccccc ccccc@{}}
\toprule
\textbf{Defense} & \textbf{Method} & \multicolumn{5}{c}{\textbf{TPR @ 0.1\% FPR ($\downarrow$)}} & \multicolumn{5}{c}{\textbf{Balanced Accuracy ($\downarrow$)}} \\
\cmidrule(lr){3-7} \cmidrule(lr){8-12}
 & & \textbf{C100} & \textbf{C10} & \textbf{CN10} & \textbf{GTS} & \textbf{OF} & \textbf{C100} & \textbf{C10} & \textbf{CN10} & \textbf{GTS} & \textbf{OF} \\
\midrule
None & SmoothGrad   & 1.40\% & 0.10\% & 0.45\% & 20.13\% & 0.10\% & 0.85 & 0.63 & 0.73 & 0.84 & 0.70  \\
None & VarGrad      & 0.20\% & 0.20\% & 0.89\% & 29.39\% & 0.00\% & 0.86 & 0.65 & 0.73 & 0.78 & 0.70  \\
None & GradCAM      & 1.00\% & 0.10\% & 0.80\% & 12.46\% & 0.10\% & 0.86 & 0.63 & 0.71 & 0.80 & 0.68  \\
None & GradCAM++    & 0.50\% & 0.20\% & 0.20\% & 28.75\% & 0.10\% & 0.79 & 0.64 & 0.72 & 0.79 & 0.69  \\
None & IG           & 0.90\% & 0.10\% & 0.10\% & 36.10\% & 0.00\% & 0.86 & 0.63 & 0.73 & 0.83 & 0.69  \\
None & SHAP         & 0.60\% & 0.20\% & 0.30\% & 29.71\% & 0.30\% & 0.85 & 0.62 & 0.73 & 0.82 & 0.69  \\
None & LIME         & 0.70\% & 0.70\% & 1.40\% & 37.06\% & 0.69\% & 0.87 & 0.66 & 0.74 & 0.80 & 0.70  \\
\midrule
DP ($\epsilon = 100)$ & SmoothGrad   & 0.00\% & 0.00\% & 0.15\% & 0.00\% & 0.20\% & 0.51 & 0.50 & 0.49 & 0.51 & 0.55  \\
DP ($\epsilon = 100)$ & VarGrad      & 0.10\% & 0.20\% & 0.35\% & 0.00\% & 0.10\% & 0.51 & 0.50 & 0.50 & 0.50 & 0.54  \\
DP ($\epsilon = 100)$ & GradCAM      & 0.10\% & 0.10\% & 0.10\% & 0.00\% & 0.10\% & 0.51 & 0.51 & 0.50 & 0.51 & 0.53  \\
DP ($\epsilon = 100)$ & GradCAM++    & 0.10\% & 0.00\% & 0.10\% & 0.00\% & 0.10\% & 0.52 & 0.52 & 0.50 & 0.52 & 0.55  \\
DP ($\epsilon = 100)$ & IG           & 0.10\% & 0.00\% & 0.15\% & 0.00\% & 0.10\% & 0.51 & 0.49 & 0.50 & 0.53 & 0.52  \\
DP ($\epsilon = 100)$ & SHAP         & 0.10\% & 0.00\% & 0.15\% & 0.00\% & 0.00\% & 0.51 & 0.50 & 0.50 & 0.53 & 0.54  \\
DP ($\epsilon = 100)$ & LIME         & 0.20\% & 0.10\% & 0.10\% & 0.01\% & 0.20\% & 0.54 & 0.51 & 0.52 & 0.54 & 0.55  \\
\midrule
Memguard & SmoothGrad   & 1.40\% & 0.10\% & 0.50\% & 19.8\% & 0.10\% & 0.85 & 0.60 & 0.73 & 0.80 & 0.70 \\
Memguard & VarGrad      & 0.10\% & 0.20\% & 0.90\% & 23.3\% & 0.00\% & 0.86 & 0.65 & 0.72 & 0.76 & 0.69 \\
Memguard & GradCAM      & 1.00\% & 0.30\% & 0.80\% & 12.7\% & 0.10\% & 0.86 & 0.62 & 0.70 & 0.77 & 0.70 \\
Memguard & GradCAM++    & 0.70\% & 0.20\% & 0.30\% & 26.5\% & 0.15\% & 0.79 & 0.64 & 0.72 & 0.79 & 0.68 \\
Memguard & IG           & 1.00\% & 0.40\% & 0.05\% & 33.23\% & 0.00\% & 0.85 & 0.62 & 0.71 & 0.77 & 0.70 \\
Memguard & SHAP         & 0.60\% & 0.10\% & 0.35\% & 23.32\% & 0.35\% & 0.85 & 0.61 & 0.73 & 0.79 & 0.68 \\
Memguard & LIME         & 0.70\% & 0.60\% & 1.50\% & 30.12\% & 0.70\% & 0.84 & 0.67 & 0.72 & 0.81 & 0.70 \\\midrule
\textbf{TIER} & \textbf{SmoothGrad} & \textbf{0.60\%} & 0.10\% & \textbf{0.05\%} & \textbf{0.15\%} & 0.30\% & \textbf{0.59} & \textbf{0.53} & \textbf{0.55} & \textbf{0.54} & \textbf{0.54} \\
\textbf{TIER} & \textbf{VarGrad}    & \textbf{0.00\%} & \textbf{0.10\%} & \textbf{0.00\%} & \textbf{0.00\%} & 0.00\% & \textbf{0.60} & \textbf{0.55} & \textbf{0.56} & \textbf{0.53} & \textbf{0.54} \\
\textbf{TIER} & \textbf{GradCAM}    & \textbf{0.20\%} & \textbf{0.00\%} & \textbf{0.00\%} & \textbf{0.00\%} & \textbf{0.00\%} & \textbf{0.58} & \textbf{0.54} & \textbf{0.55} & \textbf{0.52} & \textbf{0.55} \\
\textbf{TIER} & \textbf{GradCAM++}  & \textbf{0.10\%} & \textbf{0.10\%} & \textbf{0.05\%} & \textbf{0.10\%} & 0.10\% & \textbf{0.60} & \textbf{0.53} & \textbf{0.55} & \textbf{0.52} & \textbf{0.53} \\
\textbf{TIER} & \textbf{IG}         & \textbf{0.10\%} & \textbf{0.10\%} & \textbf{0.00\%} & \textbf{0.10\%} & 0.10\% & \textbf{0.59} & \textbf{0.54} & \textbf{0.53} & \textbf{0.52} & \textbf{0.54} \\
\textbf{TIER} & \textbf{SHAP}       & \textbf{0.40\%} & \textbf{0.10\%} & 0.30\% & \textbf{0.00\%} & \textbf{0.00\%} & \textbf{0.59} & \textbf{0.53} & \textbf{0.54} & \textbf{0.52} & \textbf{0.54} \\
\textbf{TIER} & \textbf{LIME}       & \textbf{0.50\%} & \textbf{0.10\%} & \textbf{0.35\%} & \textbf{0.00\%} & \textbf{0.00\%} & \textbf{0.59} & \textbf{0.56} & \textbf{0.56} & \textbf{0.52} & \textbf{0.53} \\
\bottomrule
\end{tabular}
\caption{Security evaluation under membership inference attacks. \textbf{Bold} values indicate the best practical defense performance. \textit{Architectural Note: Datasets C100, C10, CN10, and GTS utilize a ResNet-18 backbone, while OF (OpenForensics) utilizes a MobileNetV2 backbone to evaluate architectural generalization.} (C100: CIFAR100, C10: CIFAR10, CN10: CINIC10, GTS: GTSRB, OF: OpenForensics).}
\label{tab:attack_rate_main}
\end{table*}

\subsection{Security and Utility Trade-offs}
\noindent\textbf{RQ3: Can the TIER framework achieve a Pareto-optimal balance between membership privacy and model utility?} To evaluate the security-utility frontier of the proposed framework, we compare TIER against three baseline configurations: an undefended model (\textit{None}), Differential Privacy (DP-SGD) via Opacus~\cite{yousefpour2022opacususerfriendlydifferentialprivacy}, and the inference-time defense MemGuard~\cite{10.1145/3319535.3363201}. For DP-SGD, we utilize $\epsilon=100$. While this represents a high privacy budget, empirical testing showed that smaller $\epsilon$ values introduced excessive noise, rendering the models untrainable or the evaluation meaningless. For completeness, we note that MemGuard is excluded from the utility evaluation as it is an inference-time defense that does not alter model parameters or training dynamics, thus maintaining the baseline model's classification performance by design.

As shown in Table~\ref{tab:defense_utility}, TIER demonstrates superior utility retention compared to noise-based defenses. The undefended baseline achieves near-perfect training accuracy but exhibits significant vulnerability. In contrast, while DP-SGD collapses target accuracy (e.g., dropping CIFAR-100 Top-1 accuracy from $\textbf{48}\%$ to $\textbf{14}\%$), TIER maintains competitive performance, achieving $\textbf{42}\%$ accuracy on the same task. Notably, the preservation of Top-5 accuracy in TIER (e.g., \textbf{0.71} vs \textbf{0.77} on CIFAR-100) suggests that while the specific rank may shift slightly, the model retains its fundamental semantic understanding, unlike DP-SGD which drops to \textbf{0.34}. On the complex OpenForensics dataset, TIER even achieves a lower test loss ($\textbf{0.56}$) than the baseline ($\textbf{1.59}$), suggesting that the trajectory regularization acts as an effective form of generalization. Furthermore, the training overhead time with TIER defense remains computationally feasible.

The security advantages of TIER are quantified in Table~\ref{tab:attack_rate_main}, which reports the success rates of attribution-guided membership inference attacks. Vulnerability of the baseline model is heterogeneous across datasets: while TPR values remain close to zero for CIFAR-10/100, CINIC-10, and OpenForensics, they rise sharply on GTSRB (exceeding $\textbf{30}\%$), indicating dataset-specific susceptibility. Importantly, despite low TPRs elsewhere, the baseline consistently exhibits high Balanced Accuracy (0.70--0.87), meaning the attacker can still discriminate members from non-members far above random chance even when few positives are flagged at stringent thresholds. This contrast highlights that TPR@0.1\% FPR captures extreme-confidence attack success, whereas Balanced Accuracy reflects overall discriminative power across the evaluation set. 

TIER suppresses both signals: across all seven explanation methods, Balanced Accuracy drops to near-random levels ($\sim \textbf{0.55}$), while TPR values are reduced to negligible levels across all datasets, including GTSRB. Thus, TIER achieves privacy protection comparable to DP-SGD without catastrophic utility loss. These findings are corroborated by threshold-independent AUC scores in the Supplementary Material (Table S1). Unlike MemGuard, which leaves Balanced Accuracy largely intact because it does not alter the underlying model manifold, TIER provides an intrinsic defense by regularizing confidence-drop trajectories. By maintaining near-baseline utility (Table~\ref{tab:defense_utility}) while reducing attack success to random-guess levels (Table~\ref{tab:attack_rate_main}), TIER demonstrates a Pareto-dominant profile for privacy-preserving explainable AI.
\begin{figure*}[htbp]
    \centering
    \begin{subfigure}[b]{0.45\textwidth}
        \includegraphics[width=\linewidth]{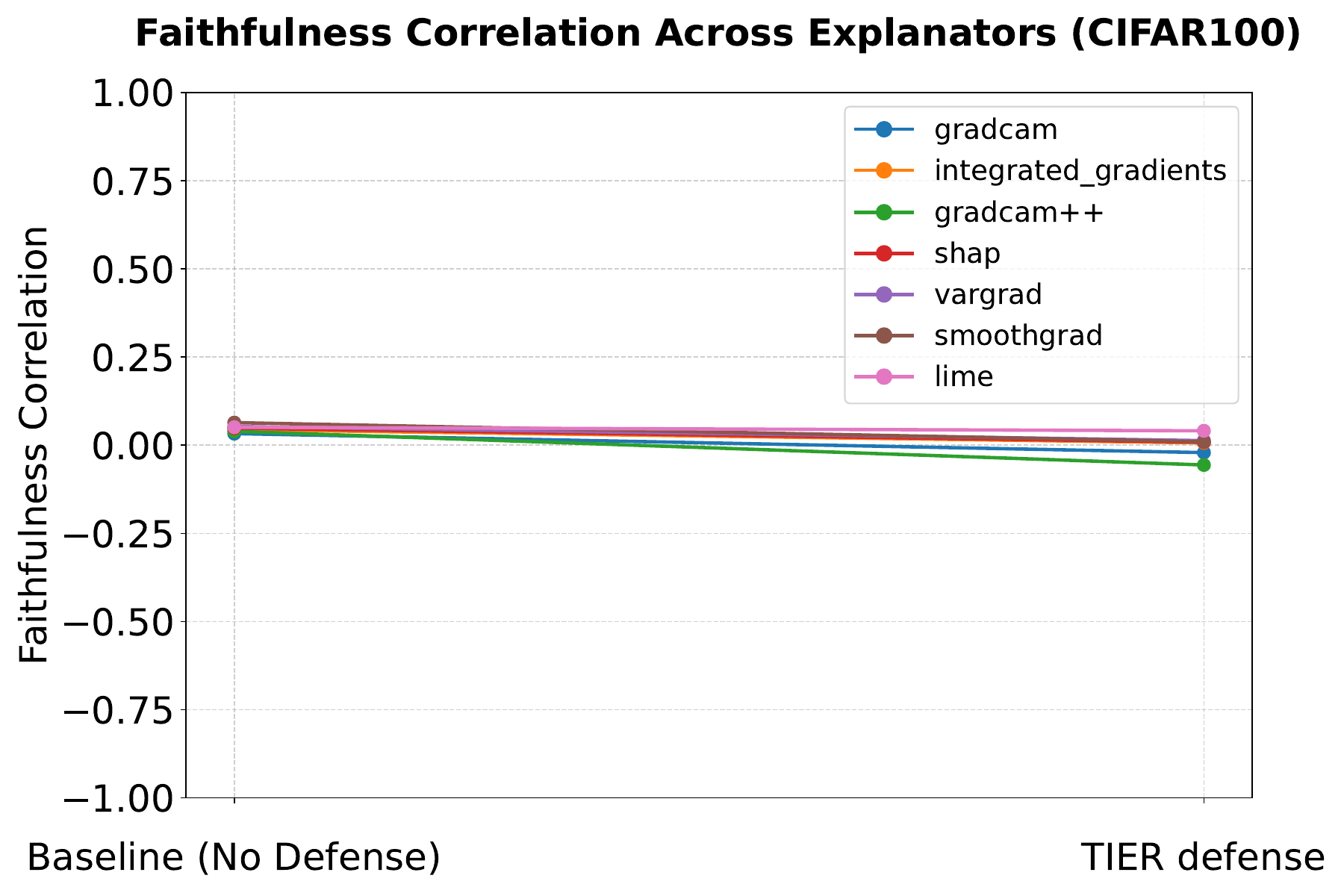}
    \end{subfigure}\hfill
    \begin{subfigure}[b]{0.45\textwidth}
        \includegraphics[width=\linewidth]{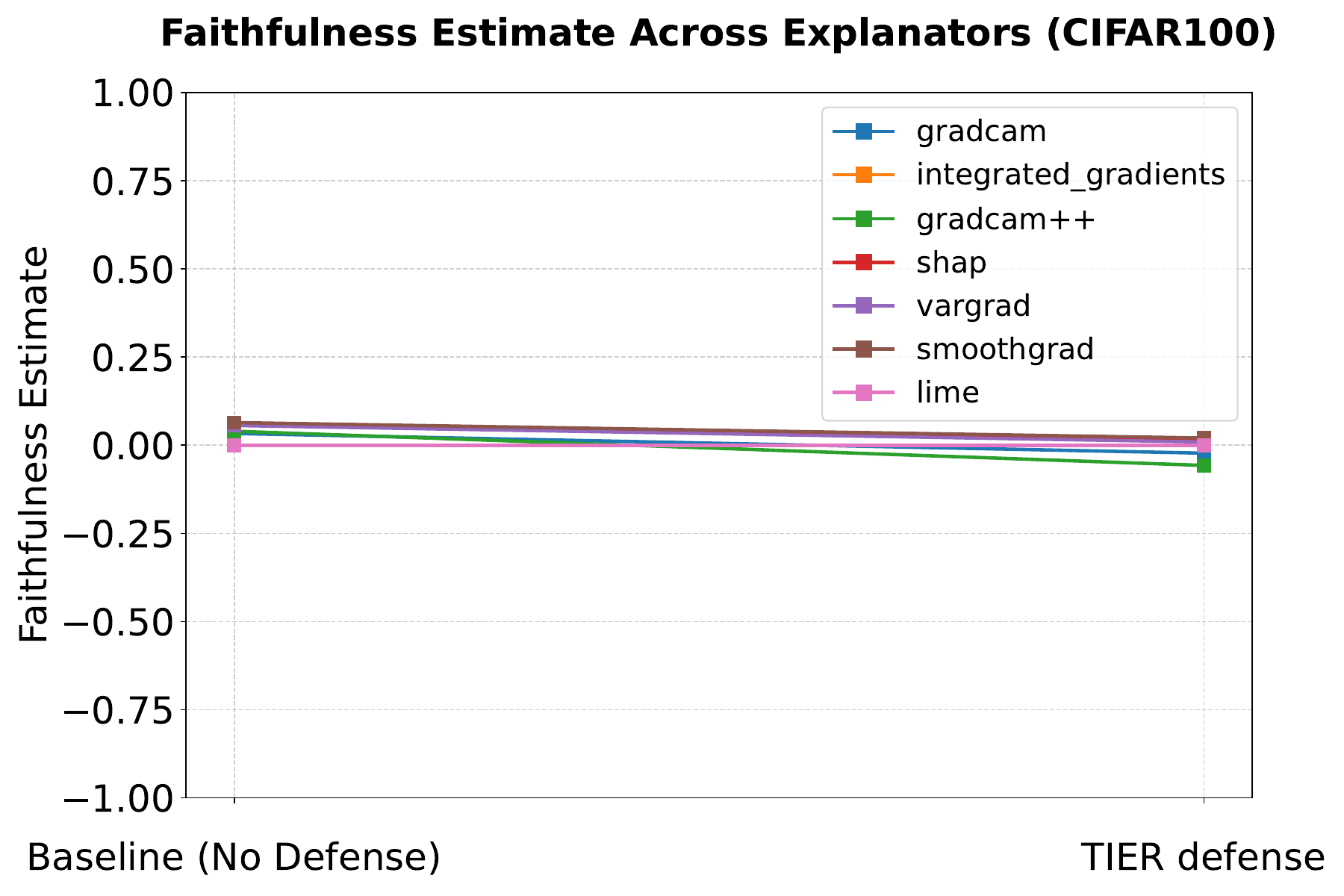}
    \end{subfigure}\hfill
    \begin{subfigure}[b]{0.45\textwidth}
        \includegraphics[width=\linewidth]{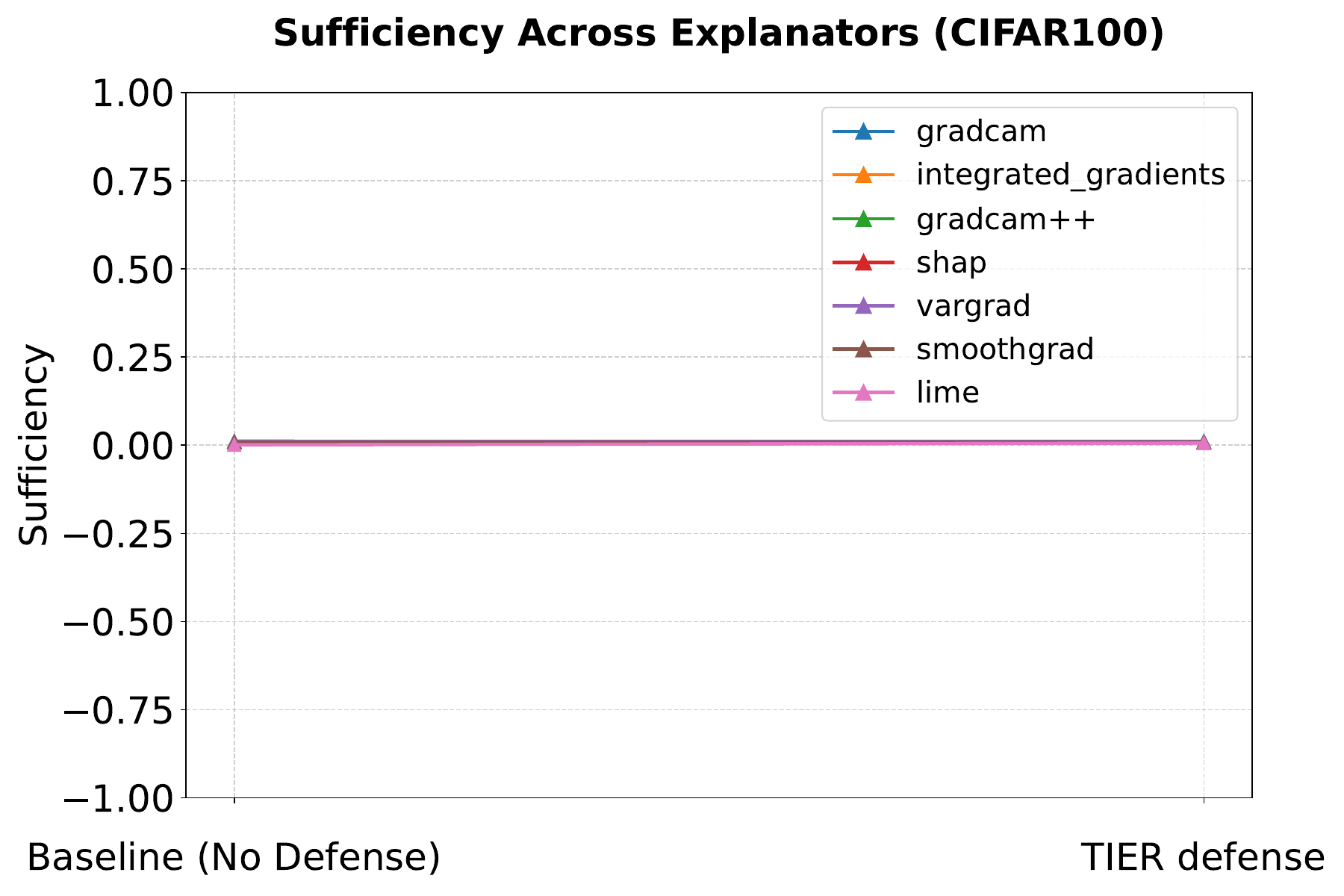}
    \end{subfigure}

    \caption{Quantitative evaluation of explanation utility on CIFAR-100. The close alignment is observed between the baseline and defended (\textit{TIER}) curves across Faithfulness and Sufficiency metrics.}
    \label{fig:main_utility_metrics}
\end{figure*}

\begin{table*}[t]
\centering
\footnotesize 
\setlength{\tabcolsep}{2.5pt} 
\begin{tabular}{ll ccccc ccccc ccccc}
\toprule
& & \multicolumn{5}{c}{\textbf{Faithfulness Corr. ($\uparrow$)}} & \multicolumn{5}{c}{\textbf{Faithfulness Est. ($\uparrow$)}} & \multicolumn{5}{c}{\textbf{Sufficiency ($\downarrow$)}} \\
\cmidrule(lr){3-7} \cmidrule(lr){8-12} \cmidrule(lr){13-17}
\textbf{Def.} & \textbf{Method} & \textbf{C100} & \textbf{C10} & \textbf{CN10} & \textbf{GTS} & \textbf{OF} & \textbf{C100} & \textbf{C10} & \textbf{CN10} & \textbf{GTS} & \textbf{OF} & \textbf{C100} & \textbf{C10} & \textbf{CN10} & \textbf{GTS} & \textbf{OF} \\ 
\midrule
\multirow{7}{*}{None} 
& SmoothGrad   & 0.06 & 0.06 & 0.04 & 0.06 & 0.03 & 0.06 & 0.07 & 0.05 & 0.06 & 0.03 & 0.01 & 0.08 & 0.07 & 0.06 & 0.40 \\
& VarGrad      & 0.05 & 0.10 & 0.03 & 0.08 & 0.00 & 0.06 & 0.09 & 0.04 & 0.08 & 0.02 & 0.01 & 0.10 & 0.09 & 0.05 & 0.45 \\
& GradCAM      & 0.03 & 0.05 & 0.02 & 0.11 & 0.00 & 0.03 & 0.05 & 0.01 & 0.04 & 0.00 & 0.01 & 0.12 & 0.11 & 0.04 & 0.51 \\
& GradCAM++    & 0.04 & 0.04 & 0.03 & 0.10 & 0.02 & 0.04 & 0.05 & 0.03 & 0.11 & 0.00 & 0.01 & 0.11 & 0.11 & 0.04 & 0.50 \\
& IG           & 0.04 & 0.06 & 0.04 & 0.05 & 0.02 & 0.06 & 0.06 & 0.05 & 0.05 & 0.03 & 0.01 & 0.08 & 0.07 & 0.04 & 0.39 \\
& SHAP         & 0.05 & 0.06 & 0.04 & 0.05 & 0.01 & 0.06 & 0.06 & 0.05 & 0.05 & 0.03 & 0.01 & 0.08 & 0.07 & 0.04 & 0.40 \\
& LIME         & 0.05 & 0.06 & 0.01 & 0.02 & 0.00 & 0.00 & 0.00 & 0.00 & 0.00 & 0.00 & 0.00 & 0.02 & 0.03 & 0.05 & 0.00 \\
\midrule
\multirow{7}{*}{\textbf{TIER}} 
& SmoothGrad   & 0.01 & -0.01 & -0.02 & \textbf{0.00} & \textbf{0.00} & 0.00 & 0.02 & -0.02 & -0.02 & -0.01 & \textbf{0.01} & \textbf{0.10} & \textbf{0.08} & \textbf{0.05} & 0.45 \\
& VarGrad      & 0.01 & -0.05 & -0.07 & -0.02 & -0.02 & 0.00 & 0.01 & -0.06 & -0.06 & -0.02 & \textbf{0.01} & \textbf{0.10} & \textbf{0.09} & \textbf{0.05} & 0.47 \\
& GradCAM      & -0.02 & \textbf{0.00} & -0.02 & -0.04 & \textbf{0.00} & -0.01 & -0.02 & \textbf{0.00} & -0.01 & -0.43 & \textbf{0.01} & \textbf{0.11} & \textbf{0.11} & 0.05 & \textbf{0.50} \\
& GradCAM++    & -0.06 & \textbf{0.04} & \textbf{0.01} & -0.06 & \textbf{0.00} & -0.00 & -0.06 & \textbf{0.04} & \textbf{0.01} & -0.07 & \textbf{0.01} & 0.12 & \textbf{0.11} & 0.05 & \textbf{0.50} \\
& IG           & 0.01 & -0.02 & -0.04 & -0.01 & \textbf{0.00} & 0.01 & 0.02 & -0.02 & -0.03 & -0.02 & \textbf{0.01} & \textbf{0.10} & \textbf{0.08} & 0.05 & 0.44 \\
& SHAP         & 0.01 & -0.03 & -0.02 & -0.01 & \textbf{0.01} & 0.01 & 0.02 & -0.02 & -0.03 & -0.02 & \textbf{0.01} & \textbf{0.09} & \textbf{0.08} & 0.05 & 0.45 \\
& LIME         & \textbf{0.04} & \textbf{0.00} & \textbf{0.00} & \textbf{0.02} & \textbf{0.00} & \textbf{0.00} & \textbf{0.00} & \textbf{0.00} & \textbf{0.00} & \textbf{0.00} & 0.01 & \textbf{0.02} & \textbf{0.03} & \textbf{0.04} & \textbf{0.00} \\
\bottomrule
\end{tabular}
\caption{Quantitative XAI metrics comparison. \textbf{Bold} values in the TIER section indicate performance that is either comparable (within $\pm 0.02$) or superior (higher for Faithfulness, lower for Sufficiency) to the undefended baseline.}
\label{tab:explanation_metrics}
\end{table*}

\subsection{Interpretability: Decoupling Faithfulness and Visual Intent}
\label{subsec:interpretability}

\textbf{RQ4: Does the TIER framework preserve the semantic interpretability and functional utility of model explanations?} 
To evaluate the privacy-interpretability tradeoff, we analyze the alignment between baseline and defended attribution maps across two distinct dimensions: \textit{Axiomatic Explanation Quality} and \textit{Visual fidelity}:
\subsubsection{\textbf{Axiomatic Quality and Feature Attribution}}
We evaluate the axiomatic properties of the explanations using the Quantus framework. As shown in Table~\ref{tab:explanation_metrics}, the introduction of TIER results in a systematic shift in \textit{Faithfulness} scores. Across all tested explainers and datasets, we observe a consistent downward trend in faithfulness correlation, with average scores shifting from low-positive baseline values ($\textbf{0.02}$ to $\textbf{0.11}$) to a compressed range near zero ($\textbf{-0.07}$ to $\textbf{0.04}$) under defense.

This global reduction is an intentional byproduct of the TIER framework; by regularizing internal gradients and trajectories to mask membership-specific information, the defense inherently decouples fine-grained pixel importance from raw model logit shifts. However, as evidenced by the Sufficiency column in Table~\ref{tab:explanation_metrics}, this does not lead to a collapse in explanation utility. In fact, TIER maintains sufficiency scores within a marginal deviation of $\pm \textbf{0.03}$ from the baseline across most configurations.

Furthermore, the stability of the defense is corroborated by the longitudinal analysis in Figure~\ref{fig:main_utility_metrics}. The close alignment between baseline and TIER-defended curves for CIFAR-100 suggests that while individual pixel-level faithfulness is traded for privacy, the aggregate functional integrity of the explanation—its ability to capture sufficient features for a prediction—remains robust.

\subsubsection{\textbf{Structural Similarity and Semantic Intent}}
The second part of our evaluation focuses on whether these internal modifications degrade the visual utility for a human user. The objective of using similarity metrics (SSIM, MSE, and Cosine) is to verify that the "semantic intent" of the attribution—identifying the correct object—is maintained.

\begin{figure*}[htbp]
    \centering
    \begin{subfigure}[b]{0.495\textwidth}
        \includegraphics[width=\linewidth]{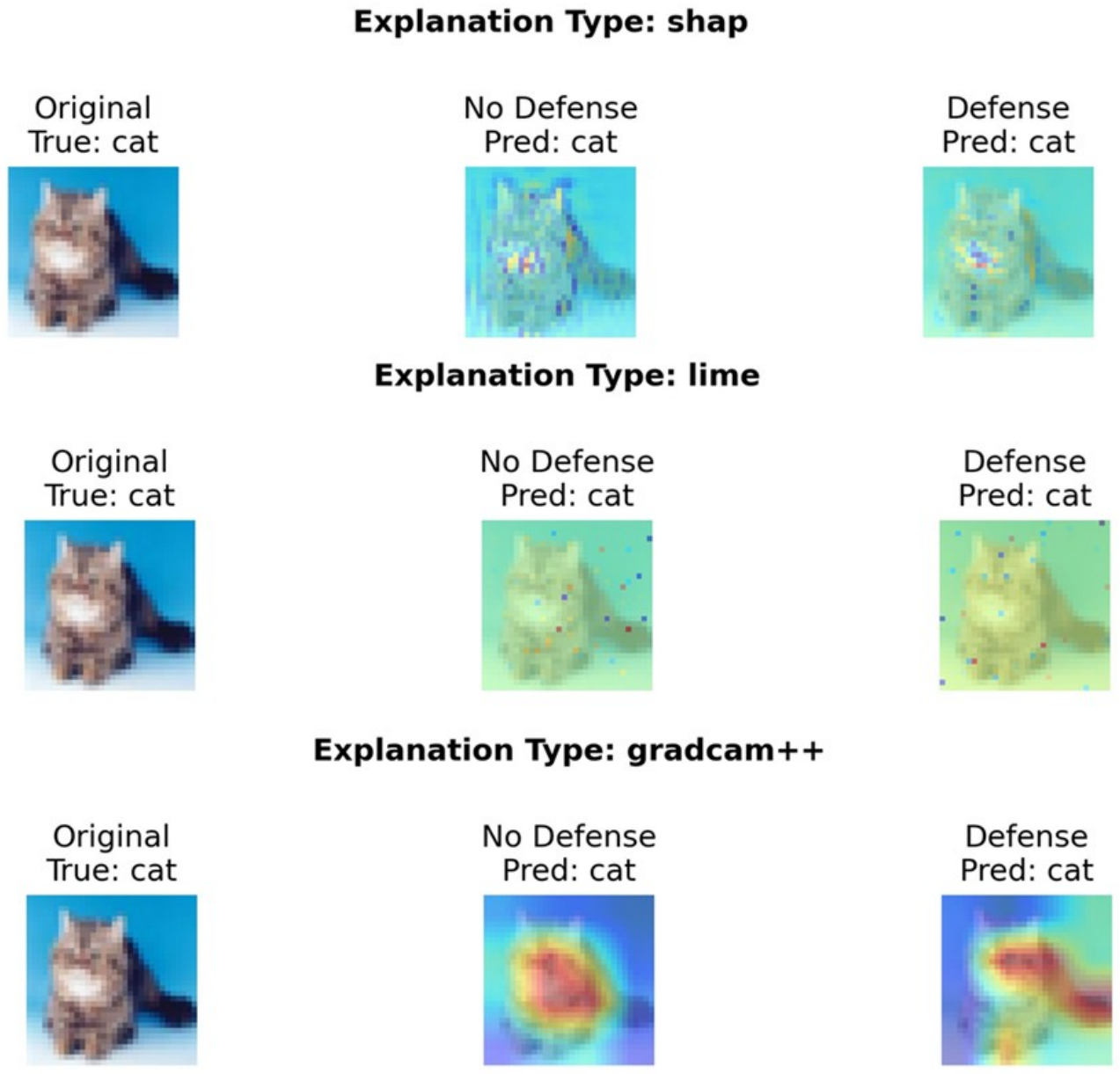}
        \caption*{CIFAR-10} 
    \end{subfigure}
    \hfill
    \begin{subfigure}[b]{0.495\textwidth}
        \includegraphics[width=\linewidth]{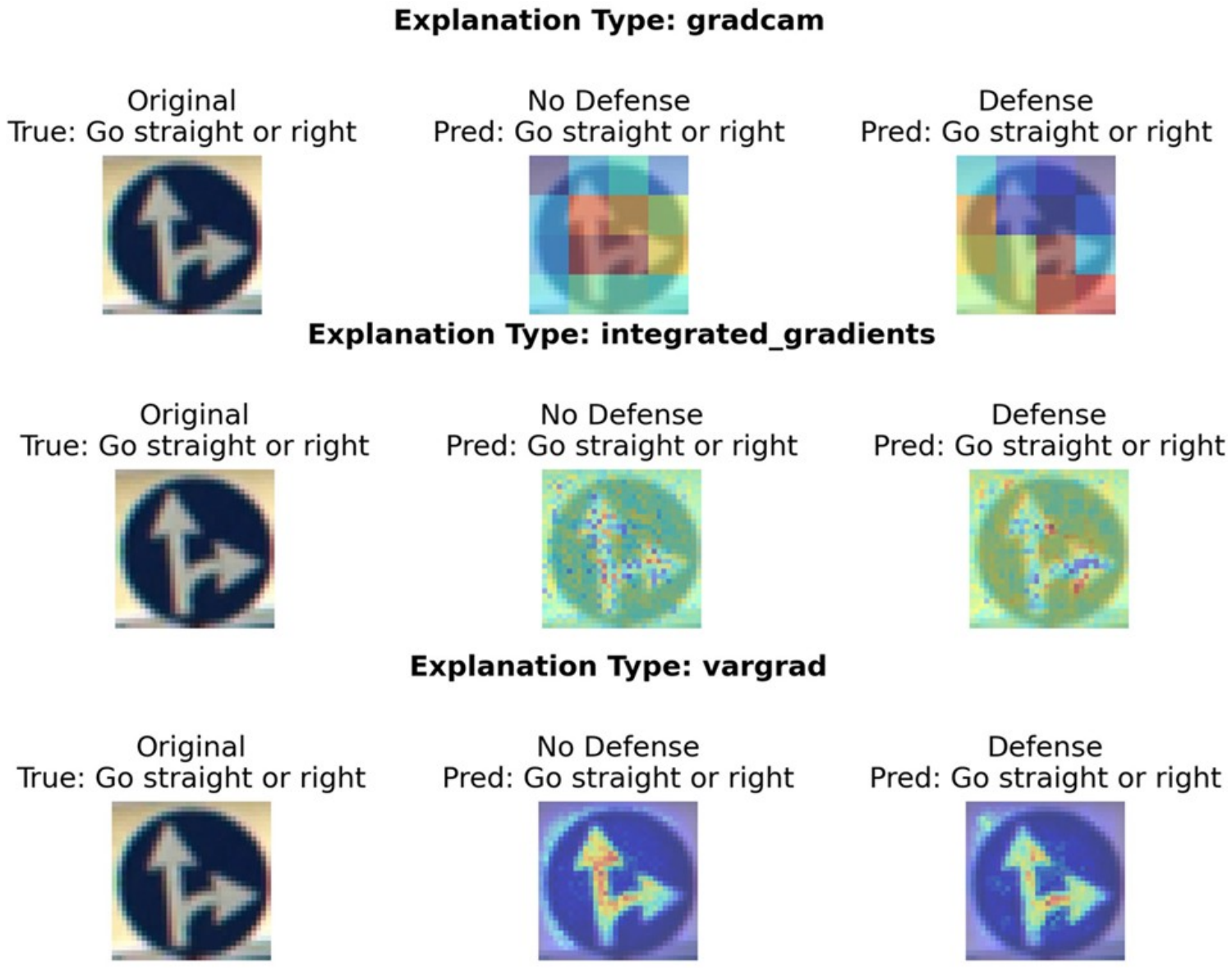}
        \caption*{GTSRB} 
    \end{subfigure}
    \caption{Qualitative comparison of attribution maps across different XAI methods and datasets. Each triplet displays the original input (left), the explanation from the undefended baseline (middle), and the explanation from the \textbf{TIER} defended model (right). The visualizations demonstrate that the defended model maintains semantic focus on primary object features (e.g., traffic sign silhouettes and animal shapes) despite the application of regularization.}
    \label{fig:main_triplets}
\end{figure*}

As summarized in Table~\ref{tab:consolidated_similarity}, the \textit{Cosine Similarity} metrics consistently exceed $0.80$ across most explainers. This indicates that the directional intent of the attribution vectors is preserved despite the privacy-preserving regularization. This is visually confirmed by the qualitative "triplets" in Figure~\ref{fig:main_triplets}, which show that the defended model continues to focus on class-relevant features (e.g., feline facial structures) even while membership information is rendered inaccessible.

\begin{table}[htbp]
\centering
\footnotesize 
\setlength{\tabcolsep}{4pt}
\begin{tabular}{llccccc}
\toprule
\textbf{Method} & \textbf{Metric} & \textbf{C100} & \textbf{C10} & \textbf{CN10} & \textbf{GTS} & \textbf{OF} \\ \midrule
\multirow{3}{*}{SmoothGrad} & SSIM $\uparrow$ & 0.21 & 0.18 & 0.23 & 0.22 & 0.13 \\
 & MSE $\downarrow$ & 0.03 & 0.03 & 0.02 & 0.02 & 0.03 \\
 & Cos $\uparrow$ & 0.96 & 0.97 & 0.97 & 0.97 & 0.97 \\ \cmidrule(lr){1-7}
\multirow{3}{*}{VarGrad} & SSIM $\uparrow$ & 0.41 & 0.32 & 0.44 & 0.44 & 0.44 \\
 & MSE $\downarrow$ & \textbf{0.02} & \textbf{0.01} & \textbf{0.01} & \textbf{0.01} & 0.03 \\
 & Cos $\uparrow$ & 0.66 & 0.54 & 0.64 & 0.64 & 0.68 \\ \cmidrule(lr){1-7}
\multirow{3}{*}{Grad-CAM} & SSIM $\uparrow$ & 0.45 & -0.10 & 0.05 & 0.05 & 0.03 \\
 & MSE $\downarrow$ & 0.09 & 0.25 & 0.18 & 0.18 & 0.20 \\
 & Cos $\uparrow$ & 0.87 & 0.62 & 0.73 & 0.73 & 0.70 \\ \cmidrule(lr){1-7}
\multirow{3}{*}{Grad-CAM++} & SSIM $\uparrow$ & \textbf{0.65} & 0.34 & 0.32 & 0.31 & 0.11 \\
 & MSE $\downarrow$ & 0.04 & 0.08 & 0.08 & 0.08 & 0.11 \\
 & Cos $\uparrow$ & 0.93 & 0.85 & 0.85 & 0.84 & 0.72 \\ \cmidrule(lr){1-7}
\multirow{3}{*}{IG / SHAP} & SSIM $\uparrow$ & 0.20 & 0.17 & 0.21 & 0.20 & 0.13 \\
 & MSE $\downarrow$ & 0.03 & 0.03 & 0.02 & 0.02 & 0.03 \\
 & Cos $\uparrow$ & 0.96 & 0.97 & 0.97 & 0.97 & 0.97 \\ \cmidrule(lr){1-7}
\multirow{3}{*}{LIME} & SSIM $\uparrow$ & 0.33 & 0.32 & 0.33 & 0.33 & 0.32 \\
 & MSE $\downarrow$ & \textbf{0.02} & 0.02 & 0.02 & 0.02 & \textbf{0.02} \\
 & Cos $\uparrow$ & \textbf{0.99} & \textbf{0.99} & \textbf{0.99} & \textbf{0.99} & \textbf{0.99} \\ \bottomrule
\end{tabular}
\caption{Similarity between defended (TIER) and original explanations. \textbf{Bold} values indicate the highest similarity/lowest error across methods. (C100: CIFAR-100, C10: CIFAR-10, CN10: CINIC-10, GTS: GTSRB, OF: OpenForensics). All metrics in this table represent the similarity between TIER-defended maps and the undefended baseline.}
\label{tab:consolidated_similarity}
\end{table}

\begin{table}[htbp]
\centering
\footnotesize 
\setlength{\tabcolsep}{5pt} 
\begin{tabular}{lccccc}
\toprule
\textbf{Dataset} & $\lambda$ & \textbf{Top-1 Acc} & \textbf{Top-5 Acc} & \textbf{Def. Acc} & \textbf{Def. AUC} \\ 
\midrule
\multirow{4}{*}{CIFAR-100}
& 0.0 & 0.41 & 0.67 & 0.71 & 0.90 \\
& 3.0 & 0.40 & 0.68 & 0.68 & 0.84 \\
& 5.0 & 0.41 & 0.71 & 0.61 & 0.78 \\
& \textbf{7.0} & \textbf{0.41} & \textbf{0.71} & \textbf{0.58} & \textbf{0.69} \\
& 11.0 & 0.39 & 0.69 & 0.54 & 0.66 \\ \midrule
\multirow{4}{*}{CIFAR-10}
& 0.0 & 0.78 & 0.97 & 0.59 & 0.68 \\
& 5.0 & 0.78 & 0.98 & 0.54 & 0.59 \\
& 7.0 & 0.79 & 0.98 & 0.54 & 0.56 \\
& \textbf{9.0} & \textbf{0.78} & \textbf{0.98} & \textbf{0.52} & \textbf{0.54} \\
& 15.0 & 0.75 & 0.97 & 0.52 & 0.52 \\ \midrule
\multirow{4}{*}{GTSRB}
& 0.0 & 0.73 & 0.91 & 0.89 & 0.96 \\
& 3.0 & 0.79 & 0.96 & 0.53 & 0.57 \\
& 5.0 & 0.77 & 0.96 & 0.53 & 0.54 \\
& \textbf{7.0} & \textbf{0.72} & \textbf{0.96} & \textbf{0.51} & \textbf{0.52} \\
& 9.0 & 0.54 & 0.89 & 0.49 & 0.49 \\ \midrule
\multirow{4}{*}{CINIC-10}
& 0.0 & 0.60 & 0.90 & 0.71 & 0.80 \\
& 5.0 & 0.62 & 0.95 & 0.58 & 0.62 \\
& 7.0 & 0.61 & 0.95 & 0.56 & 0.58 \\
& \textbf{9.0} & \textbf{0.60} & \textbf{0.95} & \textbf{0.54} & \textbf{0.57} \\
& 15.0 & 0.58 & 0.95 & 0.52 & 0.54 \\ 
\bottomrule
\end{tabular}
\caption{Sensitivity analysis of the $\lambda$ hyperparameter (with fixed $\beta=3.0$). \textbf{Bold} values indicate the selected optimal trade-off point for each dataset.}
\label{tab:lambda_tradeoffs}
\end{table}

\begin{table*}[t]
\centering
\footnotesize 
\setlength{\tabcolsep}{3.2pt} 
\begin{tabular}{ll cccc cccc cccc}
\toprule
\textbf{Dataset} & \textbf{Variant} & \multicolumn{4}{c}{\textbf{Top-1 Accuracy ($\uparrow$)}} & \multicolumn{4}{c}{\textbf{Top-5 Accuracy ($\uparrow$)}} & \multicolumn{4}{c}{\textbf{Defender Accuracy ($\downarrow$)}} \\
\cmidrule(lr){3-6} \cmidrule(lr){7-10} \cmidrule(lr){11-14}
& & $\beta$=0.0 & 0.1 & 0.3 & 0.5 & 0.0 & 0.1 & 0.3 & 0.5 & 0.0 & 0.1 & 0.3 & 0.5 \\
\midrule
\textbf{GTSRB} & Reg-only & 0.807 & \textbf{0.820} & 0.717 & 0.277 & 0.958 & 0.968 & 0.935 & 0.648 & 0.550 & \textbf{0.545} & 0.513 & \textbf{0.493} \\
($\lambda$=7.0) & Full Defense & \textbf{0.822} & 0.802 & \textbf{0.770} & \textbf{0.428} & \textbf{0.970} & \textbf{0.976} & \textbf{0.964} & \textbf{0.798} & \textbf{0.548} & 0.555 & \textbf{0.503} & 0.503 \\
\midrule
\textbf{CINIC-10} & Reg-only & \textbf{0.629} & \textbf{0.643} & \textbf{0.636} & \textbf{0.614} & 0.943 & 0.957 & \textbf{0.960} & \textbf{0.957} & \textbf{0.612} & \textbf{0.594} & \textbf{0.535} & \textbf{0.527} \\
($\lambda$=9.0) & Full Defense & 0.619 & 0.628 & 0.604 & 0.600 & \textbf{0.953} & \textbf{0.958} & 0.951 & 0.956 & 0.613 & 0.597 & 0.539 & 0.529 \\
\midrule
\textbf{C-100} & Reg-only & \textbf{0.415} & \textbf{0.421} & 0.434 & \textbf{0.403} & \textbf{0.690} & \textbf{0.702} & \textbf{0.728} & \textbf{0.713} & \textbf{0.659} & \textbf{0.663} & 0.593 & \textbf{0.540} \\
($\lambda$=7.0) & Full Defense & 0.404 & 0.410 & \textbf{0.435} & 0.396 & 0.682 & 0.698 & 0.722 & 0.701 & 0.732 & 0.674 & \textbf{0.584} & 0.552 \\
\midrule
\textbf{C-10} & Reg-only & \textbf{0.815} & \textbf{0.812} & \textbf{0.808} & \textbf{0.782} & 0.982 & \textbf{0.985} & \textbf{0.986} & \textbf{0.983} & 0.589 & \textbf{0.539} & \textbf{0.525} & 0.517 \\
($\lambda$=9.0) & Full Defense & 0.801 & 0.794 & 0.775 & 0.753 & \textbf{0.984} & 0.983 & 0.984 & 0.981 & \textbf{0.579} & 0.559 & 0.531 & \textbf{0.513} \\
\bottomrule
\end{tabular}
\caption{Ablation study comparing Reg-only and Full Defense. \textbf{Bold} values indicate superior performance in each metric category across variant comparisons.}
\label{tab:defense_comparison}
\end{table*}

\subsection{Robustness and Architectural Generalization}
\textbf{RQ5: Is the TIER defense resilient to architectural variations and hyperparameter sensitivity?} 

To verify that TIER is backbone-agnostic, we evaluated the \textbf{OpenForensics (OF)} dataset using \textbf{MobileNetV2}—a lightweight architecture common in real-world forensic settings. As shown in the OF columns of Table~\ref{tab:attack_rate_main}, the privacy trends observed on ResNet-18 are consistently reproduced. Regardless of the explainer, MIA balanced accuracy on the TIER-defended MobileNetV2 remains near random-guessing levels ($\approx \mathbf{0.54}$). The reproduction of these privacy trends on a lightweight backbone demonstrates the architectural generalizability of the TIER framework.

We further evaluate the sensitivity of the regularization coefficient $\lambda$ through the dual-perspective analysis in Table~\ref{tab:lambda_tradeoffs}. To ensure robustness against diverse attack surfaces, \textbf{``Def. Acc/AUC''} represents the success rate of a membership inference classifier trained by the defender to simulate the adversary. Detailed performance trajectories across the full range of $\lambda$ values are provided in the accuracy curves of \textbf{Fig.~S1} (Supplementary Material). 

The results in Table~\ref{tab:lambda_tradeoffs} reveal three critical dataset-specific behaviors:
\begin{itemize}
    \item \textbf{Stability:} For CIFAR-10/100, target accuracy remains robust as $\lambda$ increases, while attacker success plateaus near random levels ($\approx \textbf{52\%}$--$\textbf{58\%}$).
    \item \textbf{High Sensitivity:} GTSRB exhibits the sharpest privacy-utility trade-off; TIER successfully reduces attacker success from $\textbf{89\%}$ to $\textbf{51\%}$ at the optimal $\lambda=\textbf{7.0}$.
    \item \textbf{Decoupling:} On CINIC-10, attacker success consistently diverges from target utility, dropping to $\textbf{52\%}$ even as Top-5 accuracy stabilizes at $\textbf{95\%}$.
\end{itemize}
\noindent Based on these trends, we identify the optimal $\lambda$ values (bolded in Table~\ref{tab:lambda_tradeoffs}) that maximize inference suppression while preserving near-baseline performance.

\subsection{Ablation Study}
\label{subsec:ablation}
We conduct ablations to isolate the distinct contributions of the components described in Section~\ref{subsec:defense_components}: \textit{Reg-only} (Intrinsic Regularization) vs. \textit{Full Defense} (including Semantic Augmentation). We further evaluate the stability weight $\beta$ to identify the optimal security-utility balance (Table~\ref{tab:defense_comparison}).

The results demonstrate a clear functional synergy: while augmentations enhance model generalization, regularization is the primary driver of explanation stability and privacy. As shown in Table~\ref{tab:defense_comparison}, the \textbf{Full Defense} consistently outperforms the Reg-only variant, particularly on complex datasets like GTSRB and CIFAR-100. Increasing $\beta$ systematically reduces attacker accuracy, with the Full Defense reaching its peak privacy protection at $\beta=\textbf{0.3}$. 

Across all architectures, \textbf{$\beta = \textbf{0.3}$} emerges as the Pareto-optimal configuration, achieving near-random membership privacy while preserving competitive Top-1 and Top-5 performance. Beyond this threshold ($\beta = \textbf{0.5}$), we observe diminishing privacy gains coupled with substantial utility degradation. These findings suggest that the combination of extrinsic augmentation and intrinsic regularization provides a robust defense that reconciles high utility with rigorous privacy preservation.

\section{Summary and Takeaways}
Our analyses \textbf{(RQ1–RQ5)} show that the TIER defense balances privacy and explainability. By penalizing trajectory variance  and enforcing distributional self-consistency , it blocks membership inference attacks without relying on a specific backbone (architecture-agnostic robustness). Crucially, this privacy gain is achieved without substantial impact on model utility, as the regularization terms allow the model to maintain competitive accuracy across diverse architectures.

A key insight from fidelity analysis is the separation of \textit{visual plausibility} from {technical leakage}. Although faithfulness metrics drop, cosine similarity between baseline and defended attribution maps remains high ($>\textbf{\textbf{0.80}}$). This indicates that TIER preserves semantic structure for human interpretability (stable Sufficiency) while suppressing fine-grained gradients that leak privacy.

\section{Conclusion and Future Work}
In this research, we propose a novel defense mechanism termed Trajectory-Invariant Explanation Regularization (TIER). Our experimental results demonstrate that this approach effectively enhances model privacy—specifically by reducing the separability between member and non-member confidence-drop curves—without substantial degradation to the primary model utility. As evidenced by our implementation, TIER integrates seamlessly into standard supervised training objectives (e.g., cross-entropy loss) and necessitates no structural modifications to the underlying neural architecture. Future investigations may extend this framework to diverse data modalities, such as natural language processing and multimodal systems, or explore the development of adaptive explanation penalties tailored to counter dynamic adversarial threats.

\bibliographystyle{plain}
\bibliography{referenceBib}

@INPROCEEDINGS{10646875,
  author={Liu, Han and Wu, Yuhao and Yu, Zhiyuan and Zhang, Ning},
  booktitle={2024 IEEE Symposium on Security and Privacy (SP)}, 
  title={Please Tell Me More: Privacy Impact of Explainability through the Lens of Membership Inference Attack}, 
  year={2024},
  volume={},
  number={},
  pages={4791-4809},
  doi={10.1109/SP54263.2024.00120}}

@misc{shokri2021privacyrisksmodelexplanations,
      title={On the Privacy Risks of Model Explanations}, 
      author={Reza Shokri and Martin Strobel and Yair Zick},
      year={2021},
      eprint={1907.00164},
      archivePrefix={arXiv},
      primaryClass={cs.LG},
      url={https://arxiv.org/abs/1907.00164}, 
}

@ARTICLE{11269854,
  author={Li, Qingwen and Han, Xiao and Wang, Ruiyan and Wu, Junjie and Liu, Lanjuan},
  journal={IEEE Transactions on Dependable and Secure Computing}, 
  title={Membership Inference Attack Against Time-Series Prediction Models}, 
  year={2025},
  volume={},
  number={},
  pages={1-13},
  doi={10.1109/TDSC.2025.3637371}}

@ARTICLE{10221704,
  author={Sun, Hui and Zhu, Tianqing and Li, Jie and Ji, Shoulin and Zhou, Wanlei},
  journal={IEEE Transactions on Dependable and Secure Computing}, 
  title={Attribute-Based Membership Inference Attacks and Defenses on GANs}, 
  year={2024},
  volume={21},
  number={4},
  pages={2376-2393},
  doi={10.1109/TDSC.2023.3305591}}

@ARTICLE{11072027,
  author={Guan, Faqian and Zhu, Tianqing and Tong, Hanjin and Zhou, Wanlei},
  journal={IEEE Transactions on Dependable and Secure Computing}, 
  title={Attention-Based Membership Inference Attacks on Graph Neural Network Through Topological Features}, 
  year={2025},
  volume={22},
  number={6},
  pages={6469-6486},
  doi={10.1109/TDSC.2025.3586251}}

@ARTICLE{9793586,
  author={Liu, Lan and Wang, Yi and Liu, Gaoyang and Peng, Kai and Wang, Chen},
  journal={IEEE Transactions on Dependable and Secure Computing}, 
  title={Membership Inference Attacks Against Machine Learning Models via Prediction Sensitivity}, 
  year={2023},
  volume={20},
  number={3},
  pages={2341-2347},
  doi={10.1109/TDSC.2022.3180828}}

@misc{pawelczyk2021exploringcounterfactualexplanationslens,
      title={Exploring Counterfactual Explanations Through the Lens of Adversarial Examples: A Theoretical and Empirical Analysis}, 
      author={Martin Pawelczyk and Chirag Agarwal and Shalmali Joshi and Sohini Upadhyay and Himabindu Lakkaraju},
      year={2021},
      eprint={2106.09992},
      archivePrefix={arXiv},
      primaryClass={cs.LG},
      url={https://arxiv.org/abs/2106.09992}, 
}

@misc{duddu2022inferringsensitiveattributesmodel,
      title={Inferring Sensitive Attributes from Model Explanations}, 
      author={Vasisht Duddu and Antoine Boutet},
      year={2022},
      eprint={2208.09967},
      archivePrefix={arXiv},
      primaryClass={cs.CR},
      url={https://arxiv.org/abs/2208.09967}, 
}

@misc{zhao2022exploitingexplanationsmodelinversion,
      title={Exploiting Explanations for Model Inversion Attacks}, 
      author={Xuejun Zhao and Wencan Zhang and Xiaokui Xiao and Brian Y. Lim},
      year={2022},
      eprint={2104.12669},
      archivePrefix={arXiv},
      primaryClass={cs.CV},
      url={https://arxiv.org/abs/2104.12669}, 
}

@inproceedings{Luo_2022, series={CCS ’22},
   title={Feature Inference Attack on Shapley Values},
   url={http://dx.doi.org/10.1145/3548606.3560573},
   DOI={10.1145/3548606.3560573},
   booktitle={Proceedings of the 2022 ACM SIGSAC Conference on Computer and Communications Security},
   publisher={ACM},
   author={Luo, Xinjian and Jiang, Yangfan and Xiao, Xiaokui},
   year={2022},
   month=nov, pages={2233–2247},
   collection={CCS ’22} }

@misc{yousefpour2022opacususerfriendlydifferentialprivacy,
      title={Opacus: User-Friendly Differential Privacy Library in PyTorch}, 
      author={Ashkan Yousefpour and Igor Shilov and Alexandre Sablayrolles and Davide Testuggine and Karthik Prasad and Mani Malek and John Nguyen and Sayan Ghosh and Akash Bharadwaj and Jessica Zhao and Graham Cormode and Ilya Mironov},
      year={2022},
      eprint={2109.12298},
      archivePrefix={arXiv},
      primaryClass={cs.LG},
      url={https://arxiv.org/abs/2109.12298}, 
}

@inproceedings{10.1145/3319535.3363201,
author = {Jia, Jinyuan and Salem, Ahmed and Backes, Michael and Zhang, Yang and Gong, Neil Zhenqiang},
title = {MemGuard: Defending against Black-Box Membership Inference Attacks via Adversarial Examples},
year = {2019},
isbn = {9781450367479},
publisher = {Association for Computing Machinery},
address = {New York, NY, USA},
url = {https://doi.org/10.1145/3319535.3363201},
doi = {10.1145/3319535.3363201},
booktitle = {Proceedings of the 2019 ACM SIGSAC Conference on Computer and Communications Security},
pages = {259–274},
numpages = {16},
location = {London, United Kingdom},
series = {CCS '19}
}

@inproceedings{Abadi_2016, series={CCS’16},
   title={Deep Learning with Differential Privacy},
   url={http://dx.doi.org/10.1145/2976749.2978318},
   DOI={10.1145/2976749.2978318},
   booktitle={Proceedings of the 2016 ACM SIGSAC Conference on Computer and Communications Security},
   publisher={ACM},
   author={Abadi, Martin and Chu, Andy and Goodfellow, Ian and McMahan, H. Brendan and Mironov, Ilya and Talwar, Kunal and Zhang, Li},
   year={2016},
   month=oct, pages={308–318},
   collection={CCS’16} }

@article{Yang2025C2DPCK,
  title={C2DP: CLIP-conditioned knowledge distillation for membership inference privacy protection},
  author={Bo Yang and Hui He and Zimeng Jia and Lun Xin and Wei Yu and Zejun Wang and Renhao Lu and Weizhe Zhang},
  journal={World Wide Web},
  year={2025},
  volume={28},
  url={https://api.semanticscholar.org/CorpusID:282292845}
}

@article{articleLiu,
author = {Liu, Hao and He, Shuyao and Xu, Ting},
year = {2025},
month = {06},
pages = {},
title = {PFE-KD: A Defense Method Against Membership Inference Attack Without Loss of Accuracy},
volume = {39},
journal = {International Journal of Pattern Recognition and Artificial Intelligence},
doi = {10.1142/S0218001424500204}
}

@article{Yadav2025MitigatingBM,
  title={Mitigating black-box membership inference attack using metric mapping},
  author={Aayush Yadav and Sunil Mane},
  journal={International Journal of Information Security},
  year={2025},
  volume={24},
  url={https://api.semanticscholar.org/CorpusID:281651139}
}

@InProceedings{10.1007/978-981-96-9872-1_36,
author="Sun, ZhaoLin
and Li, YanEr
and Shi, DingYu
and Chen, Cen",
editor="Huang, De-Shuang
and Chen, Wei
and Pan, Yijie
and Chen, Haiming",
title="Ensemble Partitioning: A Defense Mechanism Against Membership Inference Attacks in ML Models",
booktitle="Advanced Intelligent Computing Technology and Applications",
year="2025",
publisher="Springer Nature Singapore",
address="Singapore",
pages="435--446",
isbn="978-981-96-9872-1"
}

@misc{CIFAR10,
  author       = {Alex Krizhevsky and Vinod Nair and Geoffrey Hinton},
  title        = {The CIFAR-10 dataset},
  year         = {2014},
  howpublished = {\url{https://www.cs.toronto.edu/~kriz/cifar.html}}
}

@misc{darlow2018cinic10imagenetcifar10,
      title={CINIC-10 is not ImageNet or CIFAR-10}, 
      author={Luke N. Darlow and Elliot J. Crowley and Antreas Antoniou and Amos J. Storkey},
      year={2018},
      eprint={1810.03505},
      archivePrefix={arXiv},
      primaryClass={cs.CV},
      url={https://arxiv.org/abs/1810.03505}, 
}

@inproceedings{Stallkamp-IJCNN-2011,
    author = {Johannes Stallkamp and Marc Schlipsing and Jan Salmen and Christian Igel},
    booktitle = {IEEE International Joint Conference on Neural Networks},
    title = {The {G}erman {T}raffic {S}ign {R}ecognition {B}enchmark: A multi-class classification competition},
    year = {2011},
    pages = {1453--1460}
}

@misc{he2015deepresiduallearningimage,
      title={Deep Residual Learning for Image Recognition}, 
      author={Kaiming He and Xiangyu Zhang and Shaoqing Ren and Jian Sun},
      year={2015},
      eprint={1512.03385},
      archivePrefix={arXiv},
      primaryClass={cs.CV},
      url={https://arxiv.org/abs/1512.03385}, 
}

@misc{sandler2019mobilenetv2invertedresidualslinear,
      title={MobileNetV2: Inverted Residuals and Linear Bottlenecks}, 
      author={Mark Sandler and Andrew Howard and Menglong Zhu and Andrey Zhmoginov and Liang-Chieh Chen},
      year={2019},
      eprint={1801.04381},
      archivePrefix={arXiv},
      primaryClass={cs.CV},
      url={https://arxiv.org/abs/1801.04381}, 
}

@inproceedings{NIPS1991_8eefcfdf,
 author = {Krogh, Anders and Hertz, John},
 booktitle = {Advances in Neural Information Processing Systems},
 editor = {J. Moody and S. Hanson and R.P. Lippmann},
 pages = {},
 publisher = {Morgan-Kaufmann},
 title = {A Simple Weight Decay Can Improve Generalization},
 url = {https://proceedings.neurips.cc/paper_files/paper/1991/file/8eefcfdf5990e441f0fb6f3fad709e21-Paper.pdf},
 volume = {4},
 year = {1991}
}

@misc{cubuk2019autoaugmentlearningaugmentationpolicies,
      title={AutoAugment: Learning Augmentation Policies from Data}, 
      author={Ekin D. Cubuk and Barret Zoph and Dandelion Mane and Vijay Vasudevan and Quoc V. Le},
      year={2019},
      eprint={1805.09501},
      archivePrefix={arXiv},
      primaryClass={cs.CV},
      url={https://arxiv.org/abs/1805.09501}, 
}

@Inproceedings{ltnghia-ICCV2021,
  Title          = {OpenForensics: Large-Scale Challenging Dataset For Multi-Face Forgery Detection And Segmentation In-The-Wild},
  Author         = {Trung-Nghia Le and Huy H. Nguyen and Junichi Yamagishi and Isao Echizen},
  BookTitle      = {International Conference on Computer Vision},
  Year           = {2021}, 
}

@article{hedstrom2023quantus,
  author  = {Anna Hedstr{\"{o}}m and Leander Weber and Daniel Krakowczyk and Dilyara Bareeva and Franz Motzkus and Wojciech Samek and Sebastian Lapuschkin and Marina Marina M.{-}C. H{\"{o}}hne},
  title   = {Quantus: An Explainable AI Toolkit for Responsible Evaluation of Neural Network Explanations and Beyond},
  journal = {Journal of Machine Learning Research},
  year    = {2023},
  volume  = {24},
  number  = {34},
  pages   = {1--11},
  url     = {http://jmlr.org/papers/v24/22-0142.html}
}

\end{document}